# Human Echolocation in Static Situations: Auditory Models of Detection Thresholds for Distance, Pitch, Loudness and Timbre

Bo N. Schenkman[1]*[¶], Vijay Kiran Gidla[2][¶],

[1]CTT - Centre for Speech Technology, Department of Speech, Music and Hearing
Royal Institute of Technology, Stockholm, Sweden.
*Corresponding author, email: bosch@kth.se
[2]Blekinge Institute of Technology, Karlskrona, Sweden.
Email: vijaykirangidla@outlook.com

[¶]The authors contributed equally to this work.

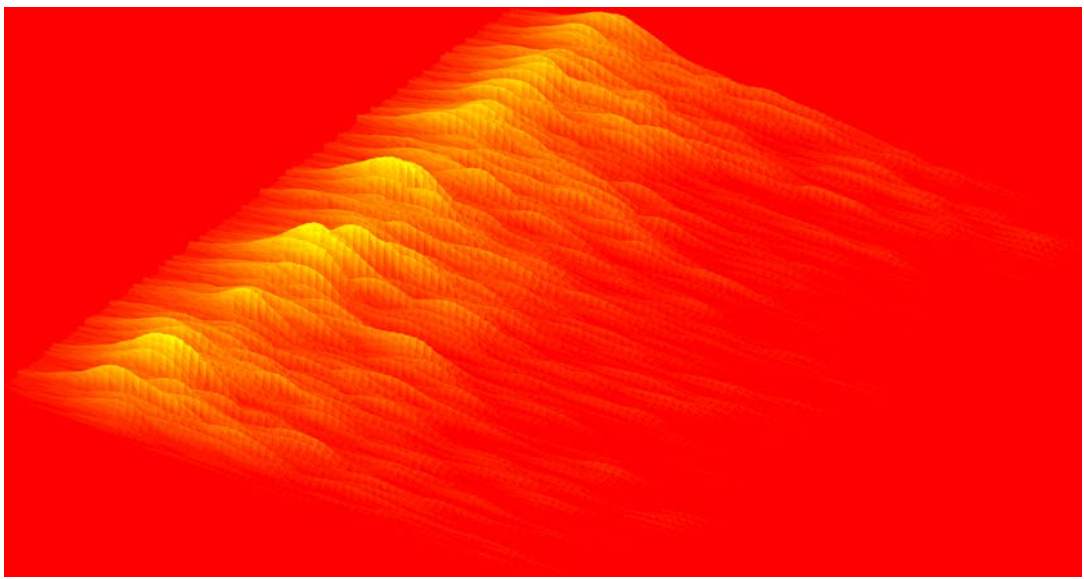

Human echolocation describes how people use reflected sounds to obtain information about their ambient world. We investigated, by using auditory models, how three perceptual parameters, loudness, pitch and sharpness, determine echolocation. We used acoustic recordings from two previous studies, both from stationary situations, and their resulting perceptual data as input to our analysis. An initial analysis was on the room acoustics of the recordings. The parameters of interest were sound pressure level, autocorrelation and spectral centroid. The auditory models were used to analyze echolocation resulting from the perceptual variables, i.e. loudness, pitch and sharpness. Relevant auditory models were chosen to simulate each variable. Based on these results, we calculated psychophysical thresholds for detecting a reflecting object with constant physical size. A non-parametric method was used to determine thresholds for distance, loudness, pitch and sharpness. Difference thresholds were calculated for the psychophysical variables, since a 2-Alternative-Forced-Choice Paradigm had originally been used. We found that (1) blind persons could detect objects at lower loudness values, lower pitch strength, different sharpness values and at further distances than sighted persons, (2) detection thresholds based on repetition pitch, loudness and sharpness varied and depended on room acoustics and type of sound stimuli, (3) repetition pitch was useful for detection at shorter distances and was determined from the peaks in the temporal profile of the autocorrelation function, (4) loudness at shorter distances provides echolocation information, (5) at longer distances, timbre aspects, such as sharpness, might be used to detect objects. We also discuss binaural information, movements and the auditory model approach. Autocorrelation was assumed as a proper measure for pitch, but the question is raised whether a mechanism based on strobe integration is a viable possibility.

# Author summary

Blind people use other senses than vision to orient and move around. Hearing, especially echolocation, i.e. reflection of echoes from objects, might be used to detect objects. Sounds can be produced by the person, by mouth or cane tapping, but can also arise from the environment, e.g. from traffic. We have been studying perceptual processes when blind people use echolocation by reanalyzing results from earlier studies. We used models of hearing to understand how these sounds are processed. Blind people could detect objects in more difficult conditions and at longer distances than sighted people. Detection was based on pitch, loudness and a timbre aspect, sharpness. The perception of pitch was useful at shorter distances and depended on the time properties of the sound and on loudness. At longer distances timbre aspects, such as sharpness, might be used to detect objects. Understanding echolocation is important for how people manage to move around in their surroundings and how our senses process spatial information of the environment. The issues involved are of interest for physiologists, psychologists and biologists. Most importantly, the issues are of practical relevance for blind people learning how to safely move around and for the design of mobility aids.

# Introduction

Persons with blindness use echolocation to obtain information about their surroundings. A person or a source in the environment emits a sound and the reflection is perceived. Both static and dynamic means are used for this sensory information. In both cases the person has to perceive if an object or obstacle is in front of him/her. This perceptual decision is determined by a threshold of detection. The threshold may vary as a function of a number of variables, like the character of the sound emitted, the rates of this sound, its position relative to the person, if motion is involved and the experience and expertise in echolocation. For a review of human echolocation, see Stoffregen and Pittenger [1], Kolarik et al [2] and Thaler and Goodale [3]. Physical properties may have different effects on psychoacoustic parameters that are used to determine if an object is in front or not. Three psychoacoustic parameters are particularly important as sources for human echolocation, viz. pitch in the form of repetition pitch, loudness and spectral information that is perceived as timbre. We describe how the information provided by pitch, loudness or timbre may result in their respective detection thresholds for echolocation. We limit ourselves to stationary situations, i.e. when neither object nor person is moving. When movement is involved, more potential information may be provided (Wilson, [4]; Wallmeier and Wiegrebe, [5]). We also determine at what distance there is a threshold when a person may detect a reflecting object. A number of auditory models were applied to the physical stimuli and we related the results of these models to the perceptual responses of participants from two previous empirical studies, Schenkman and Nilsson [6] and Schenkman, Nilsson and Grbic [7] which also will be referred to as SN2010 and SNG2016, respectively.

Psychoacoustic and neuroimaging methods are very useful for describing the high echolocating ability of the blind and their underlying processes. However, they do not fully reveal the information in the acoustic stimulus that determines echolocation (at least when the source for the information is not known) and how this information is encoded in the auditory system. We wanted to know how this information is represented and processed in the human auditory system. One fruitful way to study the information necessary for human echolocation is by signal analysis on the acoustic stimulus. However, such an analysis is only directed at the physical properties of the sound, and does not fully show how the information is represented in the human auditory system. To investigate this, we used auditory models which mimic human hearing. Analyzing the acoustic stimulus using these models provide insight into the processes for human echolocation. They may also allow testing of hypotheses by comparing models (Dau [8]).

Loudness, pitch (Schenkman and Nilsson [9]) and timbre are three perceptual attributes of an acoustic sound that are relevant for human echolocation. The backgrounds of these attributes for modeling human echolocation are discussed next.

## Loudness

Loudness is the perceptual attribute of sound intensity and is defined as that attribute of auditory sensation in terms of which sounds can be ordered on a scale from quiet to loud (ASA 1973 [10]). The dynamic range of the auditory system is wide and different mechanisms play a role in intensity discrimination. Psychophysical experiments suggest that neuron firing rates, spread of excitation and phase locking play a role in intensity perception, but the latter two may not always be essential. A disadvantage with the neuron firing rates is that, although the single neurons in the auditory nerve can be used to explain the intensity discrimination, this does not explain why the intensity discrimination is not better than observed, suggesting that the discrimination is limited by the capacity of the higher levels in the auditory system, which may also play a role in intensity discrimination (Moore 2013 [11]). Several models (Moore [11] pp 139 - 140) have been proposed to calculate the average loudness that would be perceived by listeners. The basic structure of these models is that, initially the outer and middle ear transformations are performed and then the excitation pattern is calculated. The excitation pattern is transformed into specific loudness, which involves a compressive non-linearity. The total area calculated for the specific loudness pattern is assumed to be proportional to the overall loudness. Therefore, independent of the mechanism underlying the perception of loudness, the excitation pattern is the essential information that is needed for an auditory model of loudness, and thus also for understanding human echolocation in its utilizing of loudness.

## Pitch

Pitch is "that attribute of auditory sensation in terms of which sounds may be ordered on a musical scale" (ASA 1960 [12]). One view of the underlying mechanisms of pitch is that, as the cochlea is assumed to perform a spectrum analysis, the acoustic vibrations are transformed into the spectrum, coded as a profile of discharge rate across the auditory nerve. An alternative view proposes that the cochlea transduces the acoustic vibrations into temporal patterns of neural firing. These two views are known as the place and time hypotheses. According to the place hypothesis, pitch is determined from the position of maximum excitation along the basilar membrane, within the cochlea. This explains how the pitch is perceived by pure tones at low levels, but it fails to explain the perception of pure tones at higher levels. At such levels, due to the non-linearity of the basilar membrane, the peaks become broader and tend to shift towards a lower frequency place. This should lead to a decrease in pitch, but psychophysical experiments show that the pitch is stable. Another case where the place hypothesis fails is its inability to explain the pitch of stimuli whose fundamental is absent. According to the paradox of the missing fundamental, the pitch evoked by a pure tone remains the same if we add additional tones with frequencies that are integer multiples of that of the original pure tone, i.e. harmonics. It also does not change if we then remove the original pure tone, the fundamental (De Cheveigné 2010 [13]).

Since the time hypothesis states that pitch is derived from the periodic pattern of the acoustic waveform it overcomes the problem of the missing fundamental. However, the main difficulty with the time hypothesis is that it is not easy to extract one pulse per period, in a way that is reliable and fully general. Psychoacoustic studies also show that pitch exists for sounds which are not periodic. Of interest to our subject matter, human echolocation, is an instance of such sounds, namely iterated ripple noise. It is a sound that models some of the human echolocation signals (e.g. Bilsen [14]).

In order to overcome the limitations of the place and time hypothesis two new theories have been proposed, pattern matching (De Boer [15-16]; De Cheveigné [13]), and a theory based on autocorrelation (Licklider [17]; De Cheveigné [13]). De Boer [15-16] described pattern matching such that the fundamental partial is the necessary correlate of pitch, but it may be absent if other parts of the pattern are present. In this way pattern matching supports the place hypothesis. Later Goldstein [18], Wightman [19] and Terhardt [20] described other models for pattern matching. One problem with the pattern matching theory is that it fails to account for pitch whose stimuli have no resolved harmonics.

The autocorrelation hypothesis assumes temporal processing in the auditory system. It states that, instead of detecting the peaks at regular intervals, the periodic neural pattern is processed by coincidence detector neurons that calculate the equivalent of an autocorrelation function (Licklider [17]; De Cheveigné [13]). The spike trains are delayed within the brain by various time lags (using neural delay lines) and are combined or correlated with the original. When the lag is equal to the time delay between spikes, then the correlation is high and outputs of the coincidence detectors tuned to that lag are strong. Spike trains in each frequency channel are processed independently and the results are combined into an aggregate pattern. However, De Cheveigné [13] argued that the autocorrelation hypothesis works too well: It predicts that, pitch should be equally salient for stimuli with resolved and unresolved partials, but this is not the case. An alternative to the theory based on an autocorrelation like function is the strobe temporal integration (STI) of Patterson Allerhand, and Giguere [21]. In accordance with STI the auditory image underlying the perception of pitch is obtained by using triggered, quantized, temporal integration, instead of an autocorrelation function. The STI works by finding the strobes from the neural activity pattern and integrating it over a certain period.

Thus, there is no full understanding of how pitch is perceived. Irrespective if temporal, spectral or multi mechanisms determine pitch perception, the underlying information that the auditory system uses to detect pitch has to be the excitation pattern on the basilar membrane. Hence, the excitation pattern is the crucial information that should be simulated by an auditory model for pitch perception, and thus also for human echolocation.

## Repetition pitch

Human echolocation signals consist of an original sound along with a reflected or delayed signal. Several studies have been presented to explain the pitch perception of such sounds. Bassett and Eastmond [22] examined the physical variations in the sound field close to a reflecting wall. They reported a perceived pitch caused by the interference of direct and reflected sound at different distances from the wall; the pitch value being equal to the inverse of the delay. In a similar way, Small and McClellan [23] and Bilsen [24], delayed identical pulses and found that the pitch perceived was equal to the inverse of the delay, naming it time separation pitch, and repetition pitch, respectively. When a sound and the repetition of that sound are listened to, a subjective tone is perceived with a pitch corresponding to the reciprocal value of the delay time (Bilsen and Ritsma [25]). They explained the repetition pitch phenomenon by the autocorrelation peaks or the spectral peaks. Yost [26] performed experiments using iterated ripple noise stimuli and concluded that autocorrelation was the underlying mechanism, used by the listeners to detect repetition pitch.

## Timbre

When the loudness and pitch of an acoustic sound are similar, the subjective attribute of sound which may distinguish or identify the sound is its timbre. Timbre has been defined as that attribute of an auditory sensation which enables a listener to judge that two non-identical sounds, similarly presented and having the same loudness and pitch, are dissimilar (ANSI 1994 [27]). One example is the difference between two musical instruments playing the same tone, e.g. a guitar and a piano. Timbre is a multidimensional percept and there is no single scale on which we can order timbre. To quantify timbre one approach is to consider the overall distribution of the spectral energy. Plomp and co-workers [28-29] showed that the perceptual differences between different sounds were closely related to the levels in 18 1/3 octave bands, thus relating the timbre to the relative level produced by the sound in each critical band. Hence, generally, for both speech and non-speech sounds, the timbre of steady tones is determined by their magnitude spectra, although the relative phases may play a small role [30]. When we consider time varying patterns, there are several factors that may influence the perception of timbre: (i) periodicity; (ii) variation of the envelope of the waveform; (iii) spectrum changes over time; and (iv) what the preceding and following sounds were like. Using auditory models, timbre information can be assessed by the levels in the spectral envelope and by the variations of the temporal envelope. Another way to preserve the fine grain time interval information that is necessary for timbre perception is by the strobe temporal integration (STI) method of Patterson, Allerhand, and Giguere [21].

## Signal analysis

Signal analysis and auditory models may enable us to understand the processing of sounds for persons using echolocation, since one then can consider the transmission of the acoustic sound from the source via the internal representation to the final percept of the person. The acoustic sound travels and undergoes transformation because of the room acoustics. One should therefore first understand the information that is received at the human ear. Signal analysis is useful for this purpose, as we then can analyze the characteristics of the sound, which have been transformed due to room conditions. The second step is to analyze how the desired characteristics of the acoustic sound, that contains the information, are represented in the auditory system. Here auditory models are useful. The desired information is transformed in an analogous way to how the auditory system is known to process it. Keeping track of the information from the outer ear to the central nervous system will be an important part for describing how listeners perceive sounds and explaining the differences between groups of listeners with different characteristics, e.g. visually handicapped vs sighted persons. This is the methodology which was used for this report.

To model the auditory analysis performed by the human auditory system we used the auditory image model of Patterson, Allerhand, and Giguere [21], the loudness models of Glasberg and Moore [31-32] and the sharpness model of Fastl and Zwicker [33]. Matlab was used as the implementation environment. The auditory image model has been implemented in matlab by Bleeck, Ives, and Patterson [34] and the current version is known as AIM-MAT. The loudness and the sharpness models were implemented in PsySound3 (Cabrera, Ferguson, and Schubert [35]), a GUIdriven Matlab environment for analysis of audio recordings. AIM-MAT and PsySound3 were downloaded from https://code.soundsoftware.ac.uk/projects/aimmat and http:// www.psysound.org, respectively.

## Aims and hypothesis

The aims of the present study were:
(1) To study different components of the information in the acoustic stimulus, that determines echolocation.
(2) To determine the thresholds for different components of the information in the acoustic stimulus, that are important factors for the detection distance to reflecting objects.
 (3) To find out how the acoustic information that determines the high echolocation ability of the blind is represented in the human auditory system.
More specifically, our hypotheses were:
(1) Detection thresholds based on repetition pitch, loudness and sharpness will vary and will depend on the room acoustics and type of the sound stimuli that is used.
(2) Repetition pitch is useful for detection at shorter distances and is determined from the peaks in the temporal profile of the autocorrelation function, computed on the neural activity pattern.
(3) Detection at shorter distances, based on loudness provides information for listeners.
(4) At longer distances timbre aspects, such as sharpness information might be used by listeners to detect objects.

## Structure of the report

This report is hereafter structured as follows: In the next part "Method", subtitled "Room acoustics: Basic acoustic analyses of physical parameters" we describe the recordings used in the studies by Schenkman and Nilsson, SN2010 [6] and Schenkman, Nilsson, and Grbic SNG2016 [7], since these form the data for the present report. In the same part we present a signal analysis conducted on these recordings. We describe basic room acoustic parameters of the signals that form the basis for the physical information on the room to the persons, whether objects are present or not. No consideration is here done to auditory models. In the next part, "Models", subtitled "Auditory analysis of acoustic information", we describe the auditory models, their designs and implementation. The loudness, pitch and sharpness for the recordings of SN2010 [6] and SNG2016 [7] when analyzed using the auditory models are presented in the Results, "Loudness analysis: Excitation patterns, binaural loudness, short and long term loudness", "Pitch analysis: autocorrelation with dual profiles" and "Sharpness analysis" sections, respectively. In the section thereafter, "Threshold values, absolute and difference, for echolocation in static situations based on auditory model analysis," the thresholds for object detection are presented. This is followed by Discussion and Conclusions.

# Method

## Room acoustics: Basic acoustic analyses of physical parameters
### Sound recordings used

Here we describe briefly how the sound recordings of SN2010 [6] and SNG2016 [7] were made. For more detailed descriptions, see the original articles. In SN2010 [6], the binaural sound recordings were conducted in an ordinary conference room and in an anechoic chamber using an artificial manikin. The object was a reflecting 1.5 mm thick aluminum disk with a diameter of 0.5 m. Recordings were conducted at distances of 0.5, 1, 2, 3, 4, and 5 m between microphones and the reflecting object. In addition, recordings were made with no obstacle in front of the artificial manikin. Durations of the noise signal were 500, 50, and 5 ms; the shortest corresponds perceptually to a click. The electrical signal was a white noise. However, the emitted sound was not perfectly white, because of the non-linear frequency response of the loudspeaker and the system. A loudspeaker generated the sounds, resting on the chest of the artificial manikin.

In SNG2016 [7] recordings were conducted in an ordinary lecture room. Recordings were conducted at 100 and 150 cm distances between microphones and the reflecting object. The emitted sounds were either bursts of 5 ms each, varying in rates from 1 to 64 bursts per 500 ms or a 500 ms white noise. In contrast to SN2010 [6], the sounds in SNG2016 [7] were generated by a loudspeaker placed 1 m straight behind the center of the head of the artificial manikin. The sound recording set ups can be seen in Fig 1.

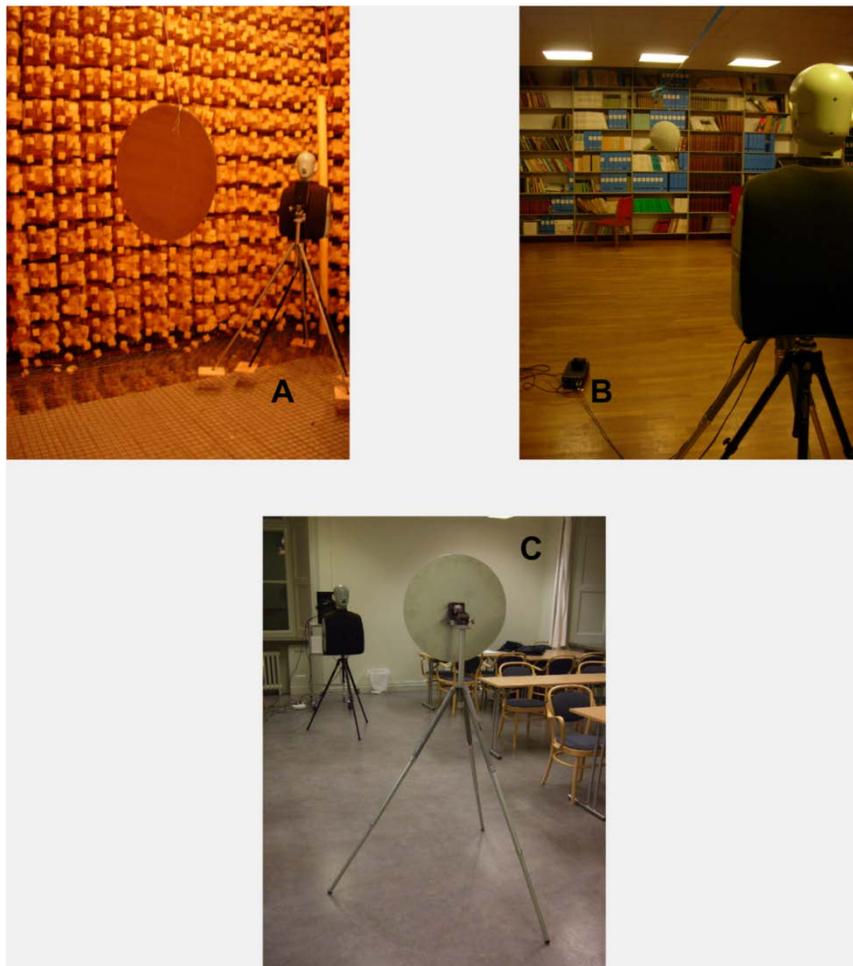

**Fig 1. Sound recordings used.** (A) anechoic room (B) conference room, with loudspeaker on the chest of the artificial manikin in Schenkman and Nilsson [6] (C) lecture room with loudspeaker behind the artificial manikin in Schenkman, Nilsson, and Grbic [7].

## Calibration of signals

Analyses were performed to determine basic room acoustical parameters relevant for human echolocation: sound pressure level, autocorrelation, and spectral centroid. Before analyzing the recordings, the recordings were calibrated by the calibrating constants (CC) using Equation (1). Based on the Sound Pressure Level (SPL) of 77, 79 and 79 dBA for the 500 ms recording without the object at the ear of the artificial manikin in the anechoic and conference rooms of SN2010 and in the lecture room of SNG2016, the CC's were calculated to be 2.4663, 2.6283 and 3.5021, respectively. A-weighting was not included in Equation (1), since the difference of using it was less than 0.5 dB and could thus be neglected. See S.1 Supporting Information for more details.

$$CC = 10^{\left(\frac{SPL - 20*log10\left(\frac{rms(signal)}{20*10^{-6}}\right)}{20}\right)} \tag{1}$$

As the recordings were binaural, both left and right ear recordings were analyzed. The recordings in SN2010 [6] and SNG2016 [7] had 10 versions of each duration and distance. For the condition with no reflecting object, there were two sets of recordings in SN2010 [6], 10 versions in each, while for the same condition in SNG2016 [7] there was one set of recordings with 10 versions. The two versions with no object in SNG2010 [6], resulted in very similar values, but are for fullness presented separately in this report. It should be noted that the recordings vary over the versions causing the term "rms(signal)" in Equation (1) to vary, thereby varying the calibrated constants for the 10 different versions. However, as the variation was very small between the versions, we decided to use only the 9th version of the 500 ms first recording without the object in SN2010 [6] and the 9th version of the 500 ms recording without the object in SNG2016 [7] to establish the calibrated constants. Another reason to choose only one version, the 9th, is that although the other versions may not have exactly the identical CC's they will be relatively calibrated with respect to the recording of version 9. For example, suppose the recording in the anechoic chamber version 1 had 67 dB SPL and version 9 had 66 dB SPL before calibration, then the levels obtained by calibrating the recordings to 77 dB SPL using the CC of the 9th version would be 78 dB SPL for version 1 and 77 dB SPL for version 9. In other words, they will give the same level difference, also after calibration.

## Sound Pressure Level (SPL)

Detection of objects by echolocation is to a certain extent based on intensity information. Hence, the SPL in dBA were calculated using Equation (2), where "RMS" is the root mean square amplitude of the signal analyzed. As has been pointed out by various authors (e.g. Rowan et al. [36]), binaural information may be utilized for echolocation purposes. We therefore calculated the SPL values for both ears. The mean SPL values of the 500 ms recordings in study SN2010 [6] and in study SNG2016 [7] are shown in Tables 1 and 2, respectively. The values for the 5 ms and 50 ms recordings are for reasons of space not shown here. As mentioned above, for the recordings with no object, two series were conducted in SN2010, each with 10 recordings. As can be seen Table 1, the values are very close to each other.

$$SPL = 20 * log10 \left( \frac{CC * rms(signal)}{20 * 10^{-6}} \right) \qquad (2)$$

**Table 1. Mean of the sound pressure levels (dBA) for the left and right ears over the 10 versions of the 500 ms duration signals in the anechoic and conference room used by Schenkman and Nilsson [6].** For the recording with no object, two series were conducted.

| Object Distance (cm) | Anechoic chamber | | Conference room | |
|:---:|:---:|:---:|:---:|:---:|
| | Left ear | Right ear | Left ear | Right ear |
| No Object, recording 1 | 77.2 | 77.9 | 79 | 78.8 |
| No Object, recording 2 | 77.6 | 77.4 | 79 | 78.8 |
| 50 | 85.2 | 88.2 | 87.5 | 87.5 |
| 100 | 81.9 | 82.6 | 82.8 | 82.4 |
| 200 | 77.1 | 78 | 79.6 | 79.5 |
| 300 | 77 | 78.2 | 78.9 | 78.9 |
| 400 | 77.1 | 78 | 79 | 78.9 |
| 500 | 77 | 78 | 79 | 78.8 |

**Table 2. Mean of the sound pressure level (dBA) for the left and right ears over the 10 versions of the 500 ms duration signals in the lecture room used by Schenkman, Nilsson and Grbic [7].**

|  | Lecture room | |
| --- | --- | --- |
| Object Distance (cm) | Left ear | Right ear |
| No Object | 79.2 | 79.6 |
| 100 | 79.6 | 81.5 |
| 150 | 79.4 | 79.7 |

The tabulated SPL values in Tables 1 and 2 show the effect of room acoustics on level differences, between the ears and between the rooms. The level differences between the recording without object and the recordings with object were in SNG2016 less than those in SN2010. This may be due to the differences in experimental setup (Fig 1) or to the acoustics of the room. The extent to which this information affected the listeners in these studies is not obvious, as loudness perceived by the human auditory system cannot be related directly to the SPL (e.g. Moore [11]). This issue is analyzed further in the section "Loudness analysis: Excitation patterns, binaural loudness, short and long term loudness".

## Autocorrelation Function (ACF)

Repetition pitch is an important aspect of how we perceive complex sounds (Bilsen [24]; Bilsen and Ritsma [25]). Schenkman and Nilsson [9] showed that this pitch, rather than loudness, is used by listeners to detect an object by echolocation. As noted above, pitch perception can often be explained by the peaks in the autocorrelation function and therefore an autocorrelation analysis was performed, which we present here.

The theoretical values for repetition pitch for the recordings of SN2010 [6] and SNG2016 [7] were calculated using Equation (3). The corresponding values for recordings with objects at distances of 50, 100, 150, 200, 300, 400 and 500 cm would be approximately 344, 172, 114, 86, 57, 43 and 34.4 Hz, assuming sound velocity to be 344 m/s. As the theory based on autocorrelation uses temporal information, repetition pitch perceived at the above frequencies can be explained by the peaks in the ACF at the inverse of the frequencies, i.e. approximately at 2.9, 5.8, 8.7, 11.6, 17.4, 23.2 and 29 ms, respectively. The autocorrelation analysis was performed using a 32 ms frame, which would cover the required pitch period. A 32 ms hop size was used to analyze the ACF for the next time instants of 64 ms, 96 ms etc. In order to compare the peaks among all the recordings, the ACF was not normalized to the limits -1 to 1.

$$RP = \frac{speed\ of\ sound}{2*distance\ of\ the\ object} \qquad (3)$$

where RP is Repetition Pitch.

In the study SN2010 [6] the participants performed well with the longer duration signals. For a single short burst the person had only one chance to perceive the signal and its echo. This can be visualized from the ACFs in Figs 2 and 3, where for the 5ms recording the peak was present only for the initial 32 ms frame. For the 500 ms recording the peak was also present for frames with time instants greater than 32 ms. (Note that for each duration of the signals an additional 450 ms silence was padded and presented to the test persons. The ACF were analyzed in the same manner, and hence the 5ms duration signal had a total duration of 455 ms and the 500 ms signal had a total duration of 950 ms.)

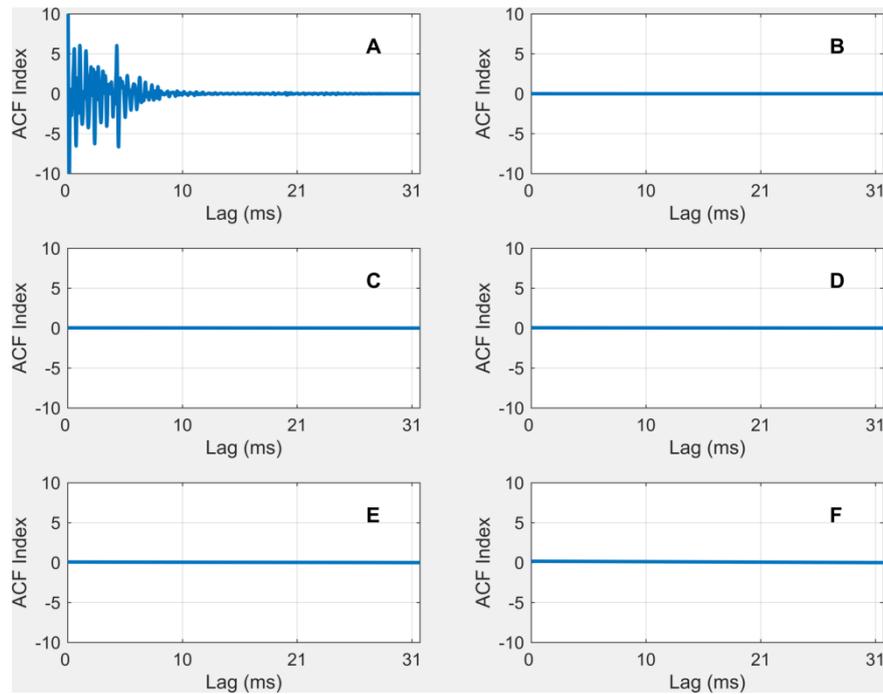

**Fig 2. The autocorrelation function of a 5 ms signal recorded in the anechoic chamber (SN2010) [6] with reflecting object at 100 cm.** The sub figures A-E show the autocorrelation function (ACF index) at 32, 64, 96, 128, 160 and 192 ms time instants of the signal (Lag), respectively. As the recording is only 5ms in duration the autocorrelation function is only present in the first 32ms frame.

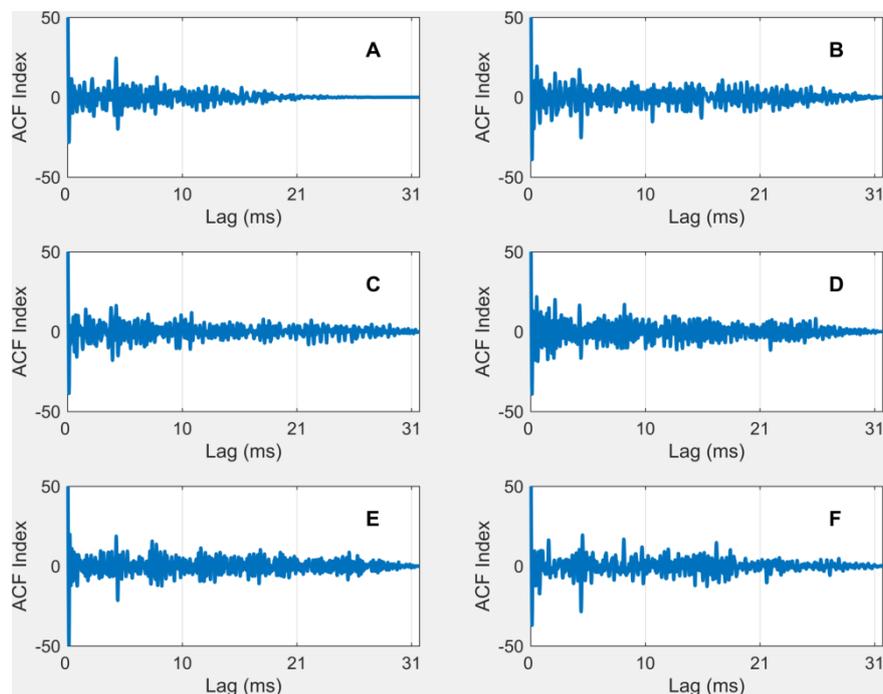

**Fig 3**. **The autocorrelation function of a 500 ms signal recorded in the anechoic chamber (SN2010) [6] with reflecting object at 100 cm.** The sub figures A-E show the autocorrelation function (ACF index) at 32, 64, 96, 128, 160 and 192 ms time instants of the signal (Lag), respectively.

Schenkman and Nilsson [6] argued that the higher detection ability of their participants for the longer duration signals may have been a result of attention, a confounding factor. Even if a test person at first was not attentive, he or she had a longer time interval available to perceive the signal. However, in Schenkman, Nilsson and Grbic [7] the performance decreased at one distance for the longer duration noise, although the repetitions were present for the frames with a time instant greater than 32 ms. These ACF functions are for reasons of space not shown here. Therefore, that longer duration signals are always beneficial for human echolocation cannot be concluded on the available results.

The peak heights at the pitch period for the recordings with object at 100 cm for the 5 ms duration signal in the conference room in SG2010 [6] were greater than those in the lecture room in SNG2016 [7]. The 500 ms duration signal with object at 100 cm in the lecture room in SN2016 [7] had a greater peak height than the 5 ms signal in the conference room in SN2010 [6], but the peak is not distinct enough when compared to the 500 ms duration signal in the conference room in SN2010 [6].

The cause for these differences in the peak heights between the two rooms, conference room in SN2010 [6] and the lecture room in SNG2016 [7], are probably the different room acoustics. The ACF depends on the spectrum of the signal, and the acoustics of the room certainly influences the peaks in the ACF. The reverberation time, T60, for the conference and the lecture room were 0.4 and 0.6 seconds, respectively. A fuller discussion of how the information carried by the peaks is represented in the auditory system is further discussed in the section "Pitch analysis: autocorrelation with dual profiles".

## Timbre: Spectral Centroid (SC)

Detection of an object could be provided by the timbre information available in a sound. Timbre is constituted by a number of different characteristics, e.g. roughness. Another characteristic of timbre perception is the spectral centroid (Peeters et al. [37]), which gives a time varying value characterizing the subjective center of the timbre for a sound. As mentioned in the Introduction, in the timbre section previously, the timbre of steady tones are mainly determined by their magnitude spectra. We believe that the spectral centroid is an important feature of human echolocation in static situations. As the recordings in SN2010 [6] and SN2016 [7] were static, the spectral centroid using the magnitude spectra of the recordings was computed to depict the timbre information.

To compute the spectral centroid, the recordings were analyzed using a 32 ms frame with a 2 ms overlap. The spectral centroid for each frame was computed by Equation (4). As the spectral centroid for each frame is a time varying function, it is plotted as a function of time. The means of the spectral centroid for the 10 versions at each condition for the 500 ms of the left ear recordings are shown in Figs 4 to 6.

$$SpectralCentroid = \frac{\sum(Frequency * FFT(frame))}{\sum(FFT(frame))} \qquad (4)$$

In SN2010 [6] for the recordings without the object, the spectral centroid was approximately below 5000 Hz. For the recordings with the object at 50 and 100 cm, the spectral centroids were approximately above 5000 Hz (Fig 4). This difference might provide information to listeners to distinguish conditions with an object from those without an object. The recordings with the object at 200 to 500 cm did not vary much when compared with the recording without the object. In SNG2016 [7] the spectral centroid was approximately 6000 Hz for all recordings (Fig 6), showing very small changes. The timbre information in SNG2016 [7] may thus at first sight not seem to be useful for echolocation. This physical analysis indicates that there was variation in the spectral centroid in the recordings of SN2010 [6] with object at shorter distances (distances shorter than 200 cm) but for longer distances the difference in the spectral centroid was almost negligible.

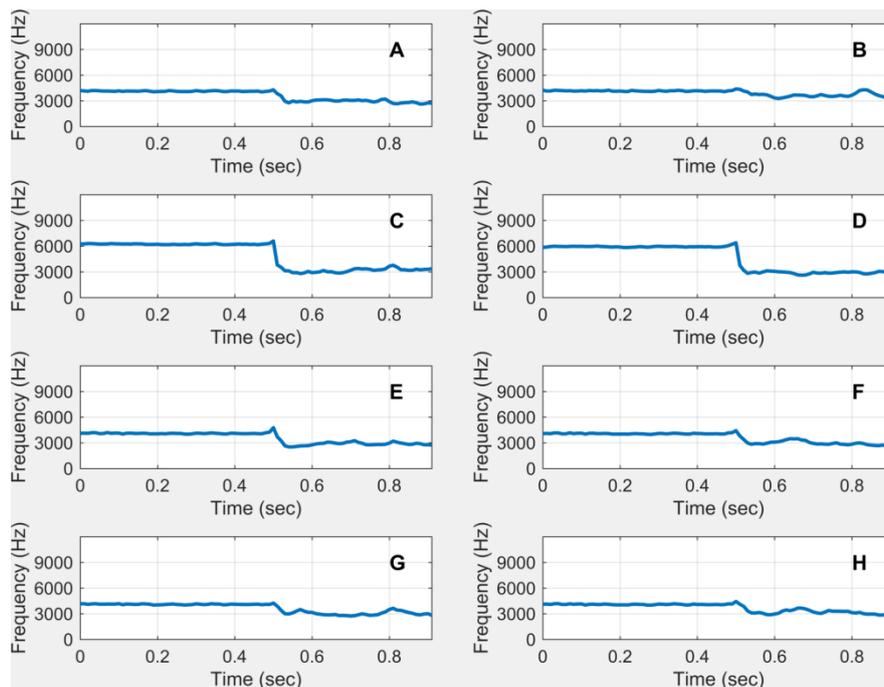

**Fig 4. The means of the spectral centroid of the 10 versions as a function of time of the left ear for a 500ms recording in the anechoic chamber** (SN2010) [6]. The sub figures A, B are for the two no object recordings and C-H are for the recordings with object at 50, 100, 200, 300, 400 and 500 cm, respectively.

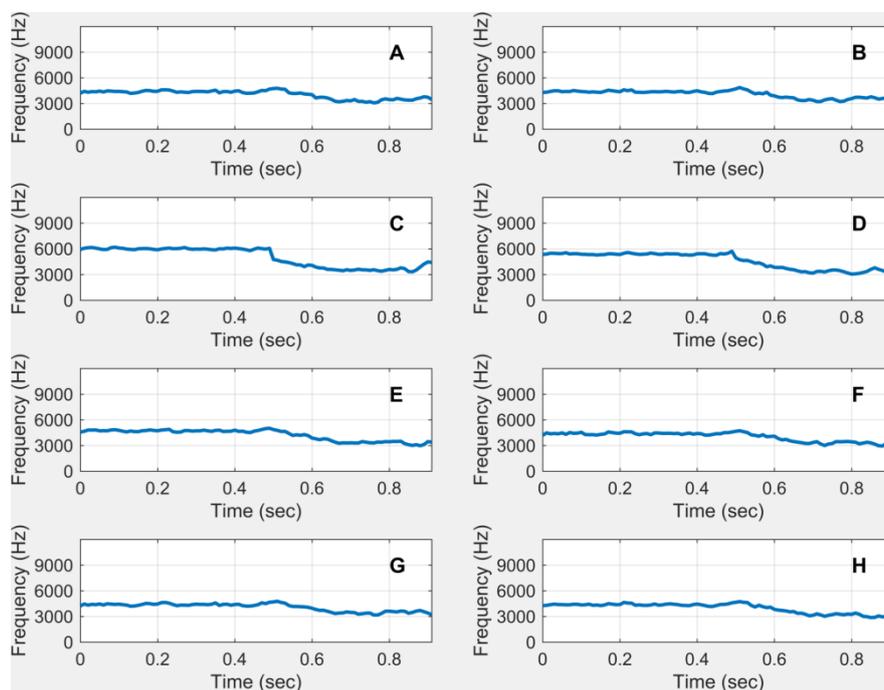

**Fig 5. The means of the spectral centroid of the 10 versions as a function of time of the left ear for a 500 ms recording in the conference room** (SN2010) [6]. The sub figures A, B are for the two no object recordings and C-H are for the recordings with object at 50, 100, 200, 300, 400 and 500 cm, respectively.

The conclusions above are based on a purely physical analysis, a Fast Fourier Transform (FFT) analysis of the sounds. However, the spectral analysis performed by the auditory system is more complex than a FFT that we have used to compute the spectral centroid. In the next section we will show that the conclusions will be modified, when we consider how human hearing works. This we do by using auditory models to analyze the sounds.

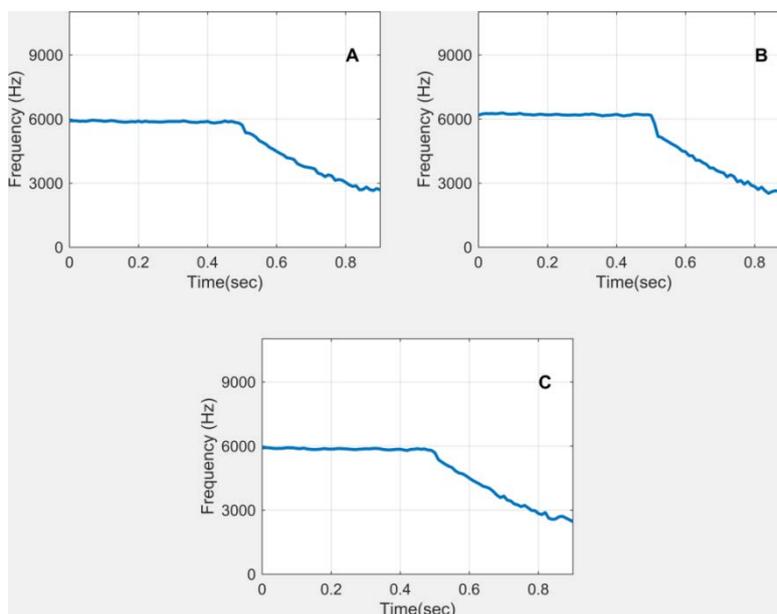

**Fig 6**. **The mean of the spectral centroid for the 10 versions as a function of time of the left ear for a 500 ms recording in the lecture room** (SNG2016) [7]. The sub figure A is for the no object recording, while sub figures B and C are for the recordings with object at 100 and 150 cm, respectively.

# Models

## Auditory analysis of acoustic information

### Premises

In the previous sections in the Method part, we presented physical parameters from the two studies SN2010 [6] and SNG2016 [7], that determine human echolocation. In this Models part we take the physical parameters from above and study how they may provide the basis for relevant auditory information to a person. In the Results part thereafter, we will connect this auditory analysis with the behavioural results.

For the auditory analysis we used the auditory image model (AIM), originally developed by Patterson et al. [38, 21] with extensions added by other authors. It is a time-domain, functional model of the signal processing performed in the auditory pathway as the system converts a sound wave into the perception that we experience when presented with a sound. This representation is referred to as an auditory image by analogy with the visual image of a scene that we experience in response to optical stimulation. The AIM simplifies the peripheral and the central auditory systems into modules. A summarily description of the AIM and how the modules were implemented in the present analysis is given below. A more detailed description of each module of AIM can be found at http://www.acousticscale. org/wiki/index.php/AIM2006_Documentation.

We used the modules described below to analyse the recordings. All the processing modules of AIM are written in matlab. The current version, as mentioned earlier, is referred to as AIM-MAT and can be downloaded from https://code.soundsoftware.ac.uk/projects/aimmat. The autocorr module was only present in the 2003 version of AIM and can be downloaded from http://w3.pdn.cam.ac. uk/groups/cnbh/aimmanual/download/downloadframeset.htm.

## First step: Pre Cochlear Processing (PCP)

The outer middle ear transformation of the acoustic sound is simulated in AIM by the Pre Cochlear Processing (PCP) module. The PCP module consists of four different Finite Impulse Response (FIR) filters, designed for different applications. These are: (i) Minimum audible field (MAF), which is suitable for signals presented in free field. (ii) Minimum audible pressure (MAP), which is suitable for systems which produce a flat frequency response. (iii) Equal loudness contour (ELC) and (iv) filter gm2002 (Glasberg and Moore [31]) are almost identical and include the factors associated with the extra internal noise at low and high frequencies. However, gm2002 uses more recent data of Glasberg and Moore [31]).

The MAF, MAP, ELC were designed using Parks-McClellan optimal equi-ripple FIR filter design algorithm, while the gm2002 was designed using a frequency sampling method. The transmission of the acoustic sound through the PCP filter can be modelled using Equation (5), where $Signal_{input}$ is the input to the AIM and $Signal_{pcp}$ is the filtered output of the corresponding PCP filter.

$$Signal_{pcp} = filter(PCP_{filter}, Signal_{input}) \qquad (5)$$

An example of the frequency response used to generate a PCP filter is shown in Fig 7.

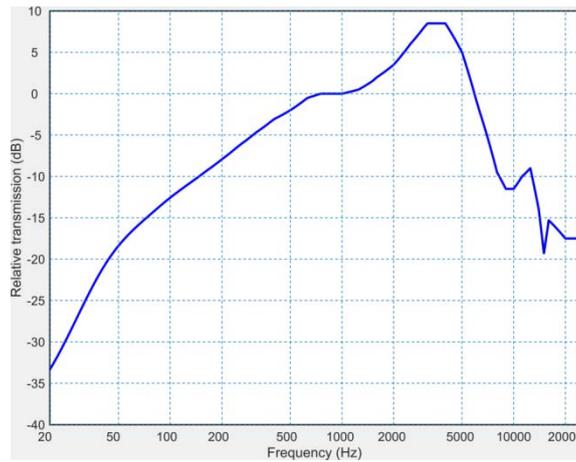

**Fig 7. The frequency response used to design the gm2002 filter of the Pre Cochlear Processing (PCP) module in the AIM.** The frequency response was obtained from the frontal field to cochlea correction data of Glasberg and Moore [31].

## Second step: Basilar Membrane Motion (BMM)

The non-linear spectral response of the basilar membrane is an important feature of the peripheral auditory system. This response is implemented in the AIM by a dynamic compressive gammachirp filter bank, dcGC (Irino and Patterson [39]). Two important properties of the Basilar Membrane Motion (BMM) are the asymmetry and the compression of the auditory filters that are made in proportion to the intensity level. These properties were designed using a compressive gammachirp filter (cGC). It is a generalized form of the gammatone filter, which was derived with operator techniques (Irino and Patterson [39]). The developments of both the gammatone and gammachirp filters are described in Patterson, Unoki, and Irino [40]. The cGC is simulated by cascading a passive gammachirp filter (pGC) with a high pass asymmetric function (HP-AF). The asymmetrical property is simulated by the pGC filter and its output is used to adjust the level dependency of the active part, i.e. the HP-AF.

Other options available for generating the BMM in AIM are the gammatone function and the pole zero filter cascade. Since the gammatone function does not depict the non-linearity of the basilar membrane, we used the default filterbank dcGC to simulate the BMM. The transformation of the BMM can be modelled using Equations (6) and (7). $Signal_{pGC}(f_c)$ is the filtered output of the pGC filterbank, fc is the centre frequency of the filter, ACF(fc) is the high pass asymmetric compensation filters and $Signal_{cGC}(f_c)$ is the final compressed output of the BMM. For a detailed description of the pGC and cGC filterbanks, see Irino and Patterson [39].

$$Signal_{pGC}(f_c) = filter\big(PGC(f_c), Signal_{pcp}\big) \qquad (6)$$

$$Signal_{cGC}(f_c) = filter\big(ACF(f_c), Signal_{pGC}(f_c)\big) \qquad (7)$$

## Third step: Neural Activity Pattern (NAP)

The basilar membrane motion is transduced into an electrical potential by the inner hair cells. The Neural Activity Pattern (NAP) is implemented in AIM by half wave rectification followed by low pass filtering. Low pass filtering is executed as phase locking is not feasible for high frequencies in the human ear.

There are three modules in the AIM to generate the NAP: (i) half wave rectification followed by compression and low pass filtering (H-C-L) (ii) half wave rectification followed by low pass filtering (H-L) (iii) two dimensional adaptive threshold (similar to H-C-L but it has adaptation which is more realistic). The choice of NAP module depends on the choice of BMM module. As noted above, we used a dcGC filter bank in our analyses, and the compression of the basilar membrane was simulated by it. The H-L module was therefore chosen to generate the NAP. This transformation can be modelled using Equation (8), where abs($Signal_{bmm}(f_c)$) is the half wave rectified signal of the basilar membrane, $f_c$ is the centre frequency of the filter, *LPF* is the low pass filter and $Signal_{nap}(f_c)$ is the modelled NAP.

$$Signal_{nap}(f_c) = filter(LPF, abs(Signal_{bmm}(f_c))) \qquad (8)$$

## Fourth step: Strobe Temporal Integration (STI)

The fourth stage in the AIM represents processing in the central nervous system. Perceptual research suggests that at least some of the fine grain time interval information is needed to preserve timbre information (Krumbholz et al [41], Patterson [42 - 43]). Auditory models often time average the NAP information, which unfortunately then loses the fine grain information. To prevent this, AIM uses a procedure called Strobe Temporal Integration (STI), which is subdivided into two modules, (i) strobe finding, and (ii) temporal integration.

***Strobe Finding (SF):*** The sub module sf2003 is used to find the strobes from the NAP. It uses an adaptive strobe threshold to issue a strobe and the time of the strobe is that associated with the peak of the NAP pulse. After the strobe is initiated the threshold initially rises along a parabolic path and then returns to the linear decay to avoid spurious strobes. The duration of the parabola is proportional to the centre frequency of the channel and the height to the height of the strobe. After the parabolic section of the adaptive threshold, its level decreases linearly to zero in 30 ms. An additional feature of sf2003 is the inter channel interaction, i.e. a strobe in one channel reduces the threshold in the neighbouring channels. An example of how the threshold varies and how the strobes are calculated is shown in Fig 8.

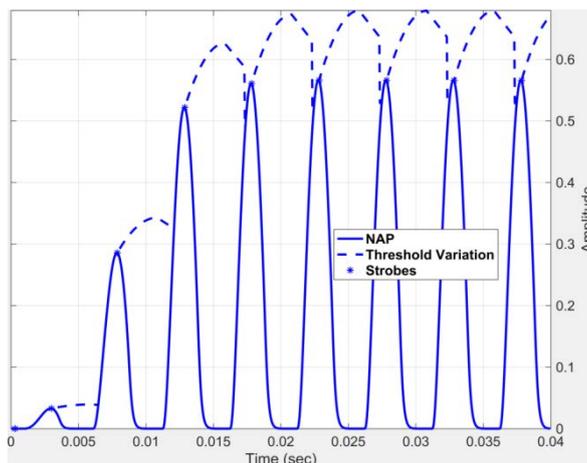

**Fig 8. The Neural Activity Pattern (NAP) of a 200 Hz pure tone in the 253 Hz frequency channel.** The dashed line shows the threshold variation and the dots indicate the calculated strobes.

**Temporal Integration (TI):** The temporal integration is implemented in AIM by a module called stabilized auditory image (SAI). The SAI in its turn uses a sub module, ti2003, to accomplish this. The ti2003 module changes the time dimension of the NAP into a time interval dimension. This works as follows: Initially, a temporal integration is initiated when a strobe is detected. If no further strobes are detected, the process continues for 35 ms and stops. If strobes are detected within the 35 ms interval, each strobe initiates a temporal integration process. To preserve the shape of the SAI to that of the NAP, ti2003 uses weighting. The new strobes are initially weighted high (the weights are also normalized so that the sum of the weights is equal to 1) making the older strobes contribute relatively less to the SAI.

### Autocorrelation Function (ACF)

As was mentioned above, the AIM offers a module, autocorr, to analyze autocorrelation processes. Corresponding physiological processes are presumed to take place in the central nervous system (e.g. Licklider [17]; Bilsen [24]; Yost [26]). By using the autocorr module one can implement models of hearing based on autocorrelation processes. The autocorr module takes the NAP as input and computes the ACF on each center frequency channel of the NAP by using a duration of 70 ms, hop time of 10ms and a maximum delay of 35ms.

# Results

## Loudness analysis: Excitation patterns, binaural loudness, short and long term loudness

In the room acoustics part earlier, for the sound pressure level analysis we described the physical intensity of a sound that may affect human echolocation. This description is necessary for understanding echolocation, but it is not sufficient. Intensity is related to the perception of loudness which is a psychological attribute, but loudness also depends on a number of other parameters, primarily but not only, frequency selectivity, bandwidth and duration of the sound. In this section we consider perceptual aspects of loudness for echolocating sounds, using the loudness model of Glasberg and Moore [31]. We chose this model instead of the AIM for the following reason.

The loudness model of Glasberg and Moore (2002) computes the frequency selectivity and compression of the basilar membrane in two stages, (1) by computing the excitation pattern and (2) by the specific loudness of the input signal. Physiologically they are interlinked and a time domain filter bank which simulates both the selectivity and the compression might be appropriate. Although there are different time domain models of the level dependent auditory filters available in AIM (e.g. dcGC), they do not give a sufficiently good fit to the equal loudness contours in ISO 2006 (Moore [44]). Since we consider this fit to be an important aspect, so instead of choosing the AIM model to model loudness, we used the model of Glasberg and Moore [31].

A loudness model should consider the outer middle ear filtering, the non-linearity of the basilar membrane and the temporal integration of the auditory system. The loudness model of Glasberg and Moore [31] estimates the loudness of steady sounds and of time varying sounds, by accounting for these features of the human auditory system. Each stage of this model is described briefly below.

## Outer middle ear transformation

The outer middle ear transformation was modelled using a finite impulse response (FIR) filter with 4097 coefficients. The response at the inner ear can be represented using Equation (9), where x and $y_{omt}$ are the signals before and after transformation, and h is the impulse response of the filter.

$$y_{omt} = filter(h, x) \tag{9}$$

## Excitation pattern

The excitation pattern is defined as the magnitude of the output of each auditory filter as a function of the filter center frequency. To compute the excitation pattern from the time domain signal, Glasberg and Moore [31] used six FFTs in parallel based on Hanning-windowed segments with durations of 2, 4, 8, 16, 32 and 64 ms, all aligned at their temporal centers. The windowed segments are zero padded, and all FFTs are based on 2048 sample points. All FFTs are updated at 1 ms intervals and each FFT was used to calculate the spectral magnitudes at specific frequency ranges. Values outside the range were discarded.

The running spectrum was the input to the auditory filters, and their output was calculated at the center frequency of 0.25 Equivalent Rectangular Bandwidth (ERB) intervals taking into account the known variation of the auditory filter shape regarding center frequency and level. The excitation pattern is defined as the output of the auditory filter as a function of center frequency (Glasberg and Moore [31]). This can be represented with Equation (10), where $E(f_c)$ is the magnitude of the output of each auditory filter with center frequency $f_c$, $Y_{omt}$ is the power spectrum of $y_{omt}$ calculated using six parallel FFT's, as mentioned above, over a 1ms interval and $W(f_c)$ is the frequency response of the auditory filter at center frequency $f_c$.

$$E(f_c) = Y_{omt} * W(f_c) \tag{10}$$

## Specific loudness (SL)

To model the non-linearity of the basilar membrane, the excitation pattern has to be converted to specific loudness. Specific loudness is the loudness in a critical band. This conversion is done in the model of Glasberg and Moore [31] using three conditions (Equation 11).

$$SL(f_c) = \begin{cases} C * \left(\frac{2E(f_c)}{E(f_c)+T_Q(f_c)}\right)^{1.5} + ((G * E(f_c) + A)^\alpha - A^\alpha) & if \ E(f_c) \leq T_Q(f_c) \\ ((G * E(f_c) + A)^\alpha - A^\alpha) & if \ 10^{10} \geq E(f_c) \geq T_Q(f_c) \\ C * \left(\frac{E(f_c)}{1.04*10^6}\right)^{0.5} & if \ E(f_c) \geq 10^{10} \end{cases}$$

(11)

$T_Q(f_c)$ is the threshold of excitation, which is frequency dependent. G represents the low level gain in the cochlear amplifier, relative to the gain at 500 Hz and above, and is frequency dependent. The parameter A is used to bring the input-output function close to linear around the absolute threshold. $\alpha$ is a compressive exponent which varies between 0.20 and 0.27. C is a constant which scales the loudness to conform to the sone scale, where the loudness of 1 kHz tone at 40 dB SPL corresponds to 1 sone and C is equal to 0.047.

Loudness depends on the intensity and bandwidth of the sound, but among other factors, also on its duration. Duration of signals is of relevance for human echolocation, and we will therefore briefly describe models of duration of loudness. The effect of the duration on the loudness was modeled by Glasberg and Moore [31] using three concepts for duration of sounds, viz. Instantaneous loudness, Short Term Loudness and Long Term Loudness. They depict the temporal integration of loudness in the auditory system and are described next.

**Instantaneous loudness (IL).** The specific loudness in each critical band has a pattern and the specific loudness over all critical bands is called specific loudness pattern. Usually the area under the specific loudness pattern is summed to give the instantaneous loudness. If the sound is binaural, then the area under the specific loudness patterns at the two ears are summed together to give the instantaneous loudness. The instantaneous loudness is an intervening variable which is used for calculations and is not a perceptual variable.

**Short Term Loudness (STL).** The Short Term loudness is determined by averaging the instantaneous loudness using an attack constant, $\alpha_a = 0.045$, and a decay constant, $\alpha_r = 0.02$ (Equation 12). The values of $\alpha_a$ and $\alpha_r$ were chosen so that the model will give reasonable predictions for variations of loudness with duration and amplitude modulated sounds (see Moore [44]).

$$STL\ (n) = \begin{cases} \propto_a * \ IL_n + (1 - \propto_a) * \ STL_{n-1} & if \ IL(n) \geq STL(n-1) \\ \propto_r * \ IL_n + (1 - \propto_r) * \ STL_{n-1} & if \ IL(n) \leq STL(n-1) \end{cases}$$

(12)

**Long Term Loudness (LTL).** The Long Term Loudness parameter was calculated by averaging the instantaneous loudness using an attack constant, $\alpha_{a1} = 0.01$ and a decay constant, $\alpha_{r1} = 0.0005$ (Equation 13). The values of $\alpha_{a1}$ and $\alpha_{r1}$ were chosen so that the model may give reasonable predictions for the overall loudness of sounds that are amplitude modulated at low rates (Moore [44]).

$$LTL\ (n) = \begin{cases} \propto_{a1} * \ STL_n + (1 - \propto_{a1}) * \ LTL_{n-1} & if \ STL(n) \geq LTL(n-1) \\ \propto_{r1} * \ STL_n + (1 - \propto_{r1}) * \ LTL_{n-1} & if \ STL(n) \leq LTL(n-1) \end{cases}$$

(13)

As noted above, loudness is also affected by binaural hearing. To model binaural loudness, a number of psychoacoustic facts have been considered (for details see Moore, 2014). Early results suggested that the level difference required for equal loudness of monaurally and diotically presented sounds was 10 dB. The subjective loudness of a sound doubles with about every 10 dB increase in physical intensity, and therefore it was assumed in the early loudness model of Glasberg and Moore [31] that loudness sums across ears. However, later results suggested that the level difference required for equal loudness is rather between 5 to 6 dB. Glasberg and Moore therefore presented a new model to account for the lower dB values based on the concept of inhibition. Inhibition occurs when a strong input in one ear lowers or even stops, i.e. inhibits, the internal response evoked by a weaker input at the other ear (Moore [44]).

Glasberg and Moore [31] implemented inhibition for binaural hearing by a gain function. Initially, the specific loudness pattern was smoothed with a Gaussian weighting function and the relative values of the smoothed function at the two ears were used to compute the gain functions of the ears. The gains were then applied to the specific loudness patterns at the two ears. The loudness for each ear was calculated by summing the specific loudness over the center frequencies and the binaural loudness was obtained by summing the loudness values across the two ears (Moore [44]). We used this procedure to calculate the binaural loudness values in this report. The binaural loudness model of Glasberg and Moore [31] has been implemented in PsySound3, a GUI-driven Matlab environment for analysis of audio recordings (http://www.psysound.org).

Glasberg and Moore [31] assumed that the loudness of a brief sound is determined by the maximum of the short term loudness, while the long term loudness may correspond to the memory for the loudness of an event that can last for several seconds. For a time varying sound (e.g. an amplitude modulated tone) it is appropriate to consider the long time loudness as a function of time to calculate the time varying loudness. However, in this report, as the stimuli presented to the participants were noise bursts and can be considered steady and brief, we follow the assumption of Glasberg and Moore [31] of using the maximum of short time loudness as a measure of the loudness of the recordings.

The means of the maxima values of Short Term Loudness in sones for the 10 versions for the 5, 50 and 500 ms recordings in the rooms of SN2010 [6] and SNG2016 [7] are presented in Tables 3, 4 and 5, respectively. From these tables, one sees that the loudness difference between the recordings without the object and with the object at 100 cm was less in the case of the lecture room of SNG2016 [7] than for the anechoic or conference room in SN2010 [6]. This may explain the low performance of the participants in the lecture room of SNG2016 [7]. The loudness values follow the same pattern as the sound pressure level analysis of the room acoustics chapter (Tables 2 and 5). However, the values in Tables 3 to 5 are psychophysical and depict not only the acoustics of the rooms but do also account for aspects of human hearing that are important for human echolocation. A comparison of the loudness results with the echolocation of persons will be made in the section "Threshold values, absolute and difference, for echolocation in static situations based on auditory model analysis", where we relate these psychophysical values to the echolocation performances.

**Table 3. Means of the maxima of Short Term Loudness, in sones, of the 10 versions for the recordings in the anechoic and conference room in SN2010 [6] and in the lecture room in SNG2016 [7] with a 5 ms duration signal.** The blank cells indicate that no recordings were made at those distances. For the SN2010 there were two series of recordings with no reflecting object.

| | Schenkman and Nilsson [6] | | Schenkman, Nilsson and Grbic [7] |
|---|---|---|---|
| Object distance (cm) | Anechoic room | Conference room | Lecture room |
| No Object, recording 1 | 13.4 | 19.3 | 15.5 |
| No Object, recording 2 | 13.3 | 19.4 | |
| 50 | 20.7 | 26.7 | |
| 100 | 20.2 | 24.4 | 17.2 |
| 150 | | | 16.2 |
| 200 | 14.4 | 21.5 | |
| 300 | 13.3 | 19.7 | |
| 400 | 13.4 | 20 | |
| 500 | 13.4 | 19.5 | |

**Table 4. Means of the maxima of Short Term Loudness, in sones, of the 10 versions for the recordings in the anechoic and conference room in SN2010 [6] with a 50 ms duration signal.** There were two series of recordings with no reflecting object.

| | Schenkman and Nilsson [6] | |
|---|---|---|
| Object distance (cm) | Anechoic room | Conference room |
| No Object, recording 1 | 40.1 | 45 |
| No Object, recording 2 | 40 | 45.1 |
| 50 | 63.7 | 69.6 |
| 100 | 52.3 | 55.7 |
| 150 | | |
| 200 | 40.3 | 47.6 |
| 300 | 40.3 | 45.1 |
| 400 | 40.2 | 45.2 |
| 500 | 40.1 | 45.0 |

**Table 5. Means of the maxima of Short Term Loudness, in sones, of the 10 versions for the recordings in the anechoic and conference room in SN2010 [6] and in the lecture room in SNG2016 [7] with a 500 ms duration signal.** The blank cells indicate that no recordings were made at those distances. For the SN2010 there were two series of recordings with no reflecting object.

| | Schenkman and Nilsson [6] | | Schenkman, Nilsson and Grbic [7] |
|---|---|---|---|
| Object distance (cm) | Anechoic room | Conference room | Lecture room |
| No Object, recording 1 | 48.1 | 52.4 | 52 |
| No Object, recording 2 | 48.1 | 52.5 | |
| 50 | 76.1 | 78.7 | |
| 100 | 62.2 | 63.6 | 54.7 |
| 150 | | | 52.5 |
| 200 | 48.4 | 54.6 | |
| 300 | 48.4 | 52.4 | |
| 400 | 48.2 | 52.6 | |
| 500 | 48.1 | 52.5 | |

# Pitch analysis: autocorrelation with dual profiles

Repetition pitch is a percept that underlies human echolocation for detecting objects. It is usually experienced as a coloration of the sound, perceived at a frequency equal to the inverse of the delay time between the sound and its reflection (Bilsen [14]; Bilsen and Ritsma [25]; see also Bassett and Eastmond [22]). As mentioned in the ACF section of Method, Room acoustics, in SN2010 and SNG2016 the reflecting objects were at distances of 50, 100, 150, 200, 300, 400 and 500 cm (although not all distances were used in both studies) resulting in delays of 2.9, 5.8, 8.7, 11.6, 17.4, 23.2 and 29 ms, where the repetition pitches would correspond to 344, 172, 114, 86, 57, 43 and 34 Hz, respectively. However, the actual delays might vary because of factors like the recording set up, speed of sound etc. and therefore the actual repetition pitch would be different. To test the presence of repetition pitch at these frequencies together with how this information would be represented in the auditory system, we used the PCP, BMM and NAP modules of the AIM, summarily presented above, to analyze the recordings from SN2010 [6] and SNG2016 [7].

The perception of repetition pitch can be created by presenting iterated rippled noise stimuli. The peaks in the autocorrelation function of these sounds are seen as the basis for repetition pith (Yost [26]; Patterson et al [45]). Hence, instead of the strobe finding and the temporal integration modules in AIM, we used the autocorr module as the final stage in our analysis to quantify repetition pitch information. Analysis by autocorrelation provides a feasible way to quantify repetition pitch, which we need to explain echolocation.

We chose not to use the strobe temporal integration as the final stage, but it does not exclude that this might be how pitch information for echolocation is represented in the auditory system. To determine whether it is autocorrelation or strobe temporal integration that better explains repetition pitch perception and possibly also physiological processes involved in the auditory system, further experiments and analysis are needed. For the interested reader, we present as an example, some results obtained using the strobe temporal integration module for a 500 ms signal, see S.2 Supporting Information.

After generating the ACF with the autocorr module in AIM, it has a dual profile development module, which sums up the ACF along both the temporal and the spectral domain. These features are relevant for human hearing in depicting how temporal and spectral information might be represented and are useful for analyzing repetition pitch. We therefore used this module to analyze the temporal and spectral results. The dual profile module plots in a single plot both the temporal and the spectral sum on frequency axis. The temporal profile and the spectral profile were scaled for this, and the inverse relation of time versus frequency (f = 1/t) was used to plot both time and frequency on a frequency scale. As an example, the dual plot for a 200 Hz pure tone is shown in Fig 9.

All the recordings were not analyzed in this way. The recordings with the object at 300 to 500 cm in study SN2010 [6] and with 2, 4, 8, 16, 32 and 64, 5 ms clicks in study SNG2016 [7] do not provide any additional information for the module and were therefore not included in this autocorrelation analysis.

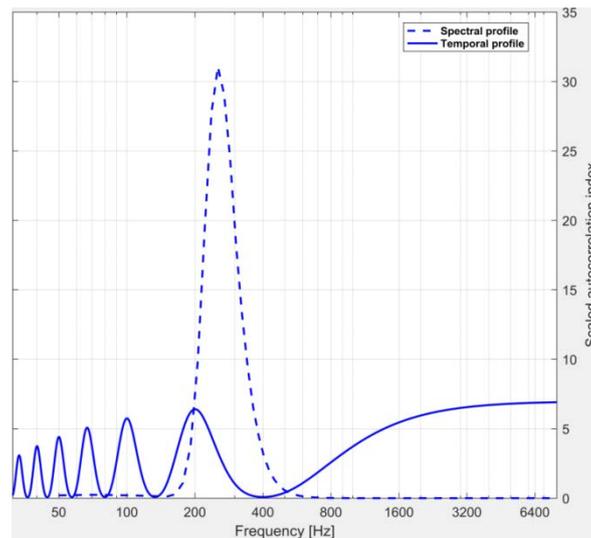

**Fig 9. The autocorrelation function of the autcorr module in the AIM for a 200Hz pure tone, where the sum of the temporal axis (spectral profile, dashed line) and the frequency axis (temporal profile, unbroken line) can be seen.** The peak in the temporal profile is at 200Hz. The peak in the spectral profile is approximately above 200Hz, this is due the high pass asymmetric function used in the dynamic compressive gammachirp filter, whose centre frequency shifts up as stimulus level increases.

The temporal profile (unbroken line in the figures below) was calculated by summing the ACF output along 100 critical bands (50Hz to 8000Hz) at each time delay. The spectral profile (dashed line in the figures below) was calculated by summing the ACF output in each critical band along a 35 ms time delay. Therefore, the temporal profile consists of the sum of 100 critical bands at every sample of 35 ms time delay, and the spectral profile consists of the sum of 35 ms delay samples at every critical band.

In the two studies that we analyze, the recordings had been presented to the participants with durations of 5, 50 or 500 ms plus an additional 450 ms of silence. Therefore, we had the same presentation duration, i.e. the whole signal was analyzed. For example, a 5 ms recording had 5 ms duration plus 450 ms of silence. However, when presenting the figures graphically, we use the first 70 ms time interval of the recordings.

The dual profiles are presented for the recordings for the three signal durations and the two rooms in study SN2010 [6] and for the 5 ms and 500 ms signal in study SNG2016 [7]. It is important to note that the amplitude scale of the y-axis is different in each sub figure of each figure. The investigated attribute here is pitch and each sub figure with reflecting object should be compared with the sub figure with No object for each condition. A distinct peak in a sub figure which is absent in the sub figure with No object indicates the potential occurrence of the perception of a pitch. One should remember that the visual impression of a peak in a sub figure with No object may misleadingly indicate an auditory peak, unless one observes the different scales on the different y-axis for the different sub figures. The next section will deal with how to select peaks based on their peak strength.

As mentioned above, the theoretical frequency of the repetition pitch for recordings with object at 100, 150 and 200 cm is 172, 114 and 86 Hz. The analysis of the 5 ms recordings show peaks approximately at these frequencies. For example, Fig 10c, 11c, 12b had peaks (marked by arrows in the subfigures) approximately at 172 Hz; Fig10d, 11d had peaks (marked by arrows in the subfigures) approximately at 82 Hz; Fig12c had a peak (marked by an arrow in the subfigure) approximately at 114 Hz. One reason that the peaks were not exactly at the theoretical values is probably due to the experimental setups of SN2016 and SNG2017 and of the room acoustics. Fig10b, 11b had no peaks at their corresponding theoretical frequencies. However, this is due to a wider range in the y axis scale (cf Fig 10, Fig11 y axis labels). The spectral profiles on the other hand did not have peaks any closer to the theoretical frequencies (cf Fig10-12 dashed lines). There were small spectral differences but these may provide timbre information but not pitch information.

In the 50 ms and 500 ms signal recordings, distinct peaks that could account for pitch perception were absent in the spectral profiles (Figs 13 to 17). We conclude that the spectral profiles (dashed line) did not provide information for pitch perception. The temporal profiles in Figs 13 to 17 might have some peaks approximately at the theoretical frequencies of the repetition pitch, but they were not clearly visible in these figures due to the scaling of the figures. Therefore, it is too premature to conclude that the temporal profile (unbroken line) is necessary for detecting the objects, if based on repetition pitch. A further analysis was therefore needed which quantified the peaks in the temporal profile.

To determine the role of temporal information for detecting objects based on repetition pitch, the pitch strength development module of AIM was used. It measures the pitch perceived based on the peak strength. We elaborate this in the next section. The temporal profiles will be shown to have peaks at the theoretical frequencies of repetition pitch which, we believe, explains the perception of repetition pitch and thus also a major cause for detection by echolocation of the reflecting objects in the two studies.

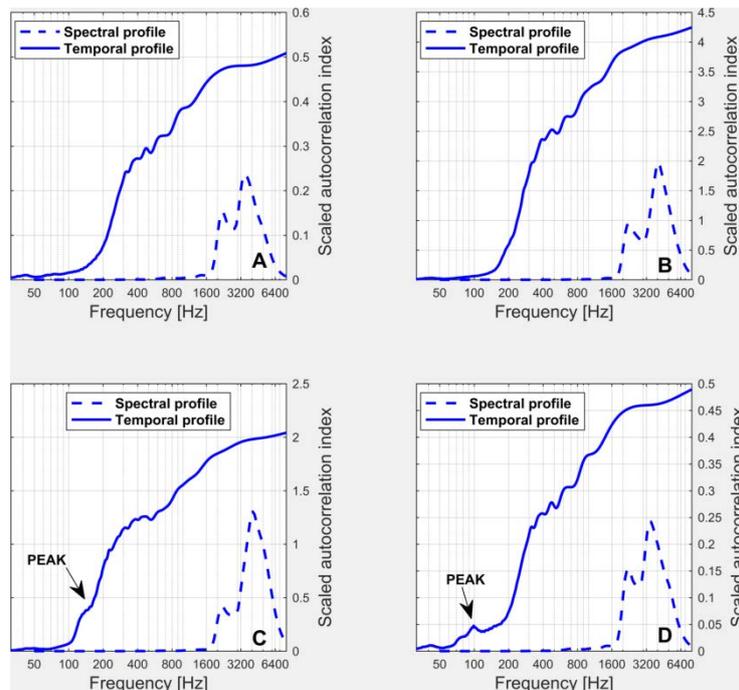

**Fig 10. The dual profile of a 5 ms signal recorded in the anechoic room** (in SN2010 [6]). The solid line is the sum of the ACF along the spectral axis and the dashed line is the sum of the ACF along the time delay axis for 70 ms time interval. The temporal and spectral profiles are scaled to be compared to each other. The 'x' axis is changed in the temporal profile by using the inverse relationship of time to frequency, f=1/t. The sub figure A is for the no object recording and sub figures B, C and D are for the recordings with object at 50, 100 and 200 cm respectively.

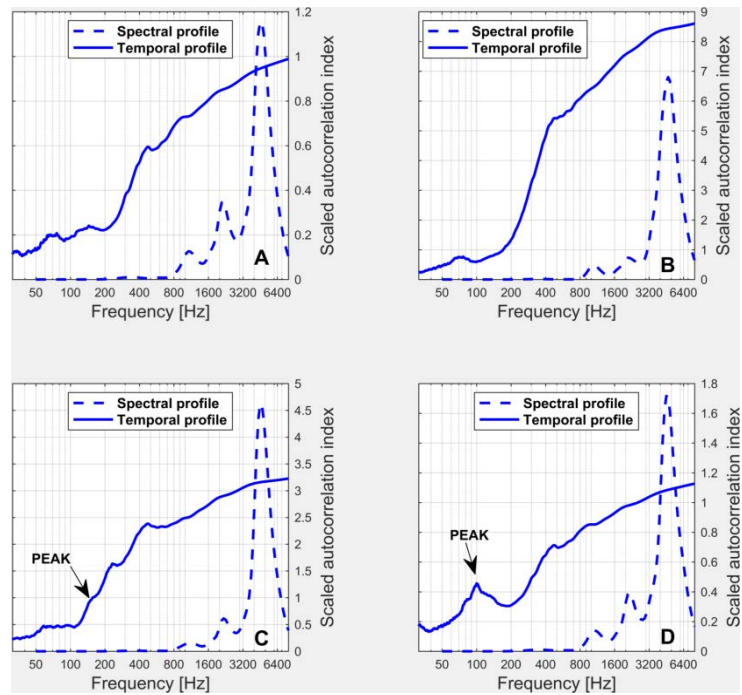

**Fig 11. The dual profile of a 5 ms signal recorded in the conference room** (in SN2010 [6]). The solid line is the sum of the ACF along the spectral axis and the dashed line is the sum of the ACF along the time delay axis for 70 ms time interval. The temporal and spectral profiles are scaled to be compared to each other. The 'x' axis is changed in the temporal profile by using the inverse relationship of time to frequency, f=1/t. The sub figure A is for the no object recording and sub figures B, C and D are for the recordings with object at 50, 100 and 200 cm respectively.

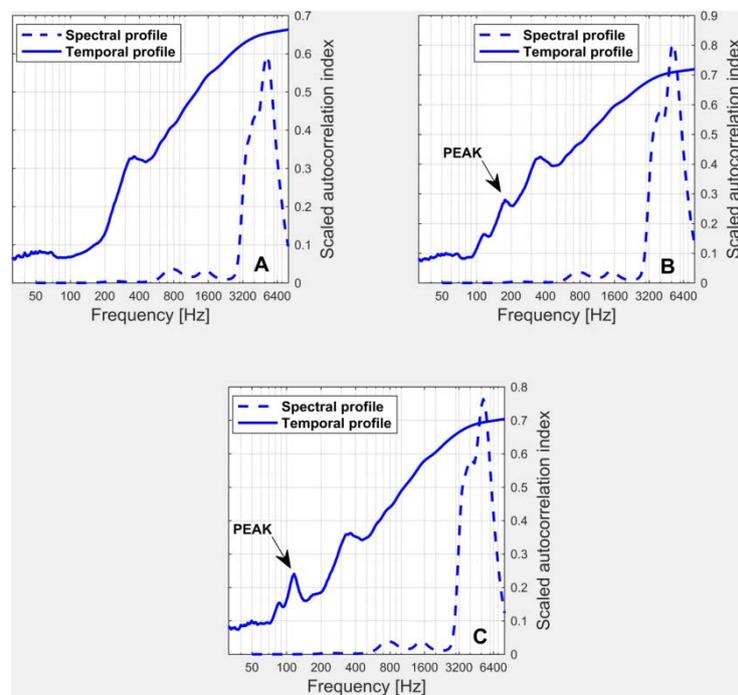

**Fig 12. The dual profile of a 5ms signal recorded in the lecture room** (SNG2016 [7]). The solid line is the sum of the ACF along the spectral axis and the dashed line is the sum of the ACF along the time delay axis for 70 ms time interval. The temporal and spectral profiles are scaled to be compared to each other. The 'x' axis is changed in the temporal profile by using

the inverse relationship of time to frequency, f=1/t. The sub figure A is for the no object recording and sub figures B and C are for the recordings with object at 100 and150 cm respectively.

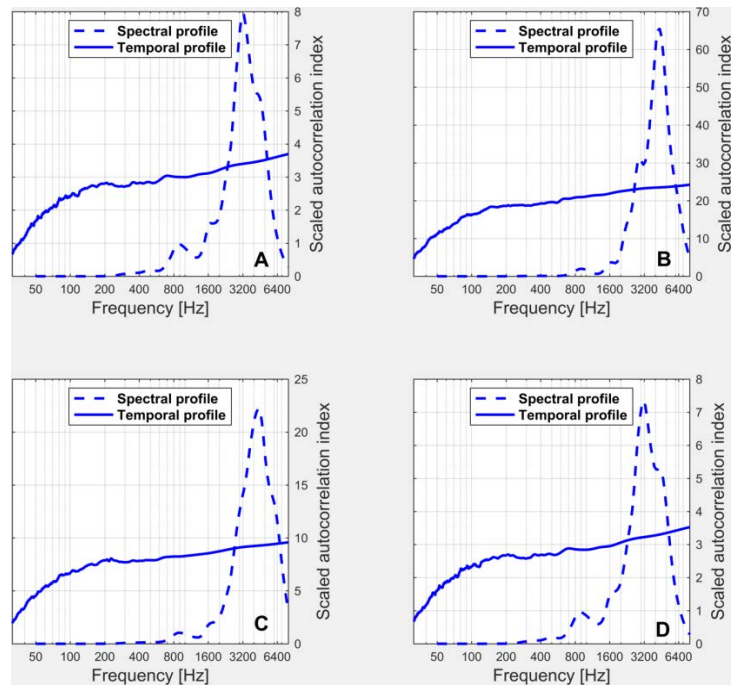

**Fig 13. The dual profile of a 50 ms signal recorded in the anechoic room** (SN2010 [6]). The solid line is the sum of the ACF along the spectral axis and the dashed line is the sum of the ACF along the time delay axis for 70 ms time interval. The temporal and spectral profiles are scaled to be compared to each other. The 'x' axis is changed in the temporal profile by using the inverse relationship of time to frequency, f=1/t. The sub figure A is for the no object recording and sub figures B, C and D are for the recordings with object at 50, 100 and 200 cm respectively.

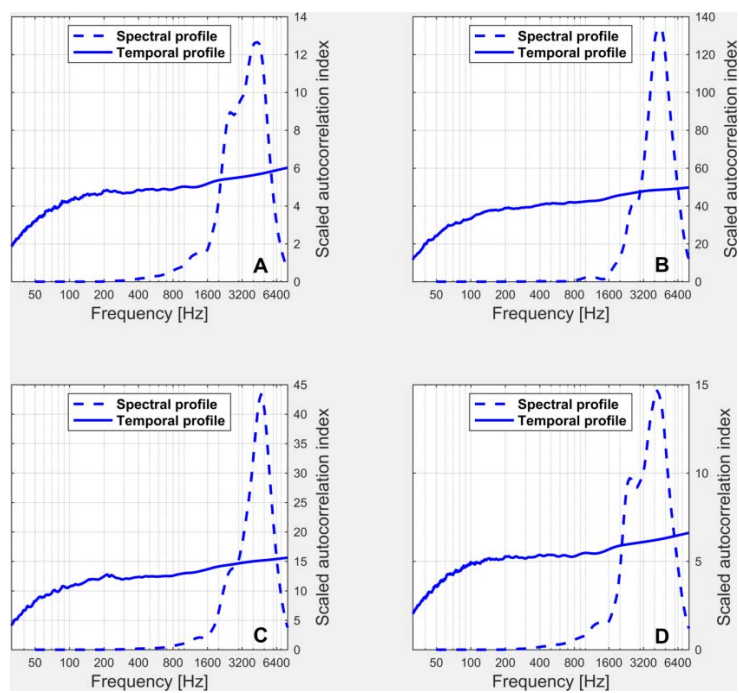

**Fig 14. The dual profile of a 50 ms signal recorded in the conference room** (SN2010 [6]).
The solid line is the sum of the ACF along the spectral axis and the dashed line is the sum of
the ACF along the time delay axis for 70 ms time interval. The temporal and spectral profiles
are scaled to be compared to each other. The 'x' axis is changed in the temporal profile by
using the inverse relationship of time to frequency, f=1/t. The sub figure A is for the no object
recording and sub figures B, C and D are for the recordings with object at 50, 100 and 200 cm
respectively.

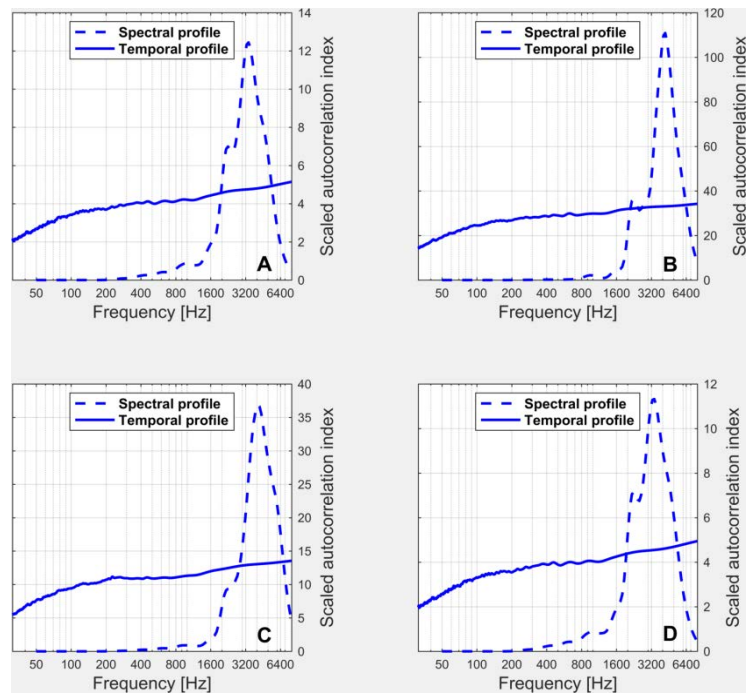

**Fig 15. The dual profile of a 500 ms signal recorded in the anechoic room** (SN2010 [6]).
The solid line is the sum of the ACF along the spectral axis and the dashed line is the sum of
the ACF along the time delay axis for 70 ms time interval. The temporal and spectral profiles
are scaled to be compared to each other. The 'x' axis is changed in the temporal profile by
using the inverse relationship of time to frequency, f=1/t. The sub figure A is for the no object
recording and sub figures B, C and D are for the recordings with object at 50, 100 and 200 cm
respectively.

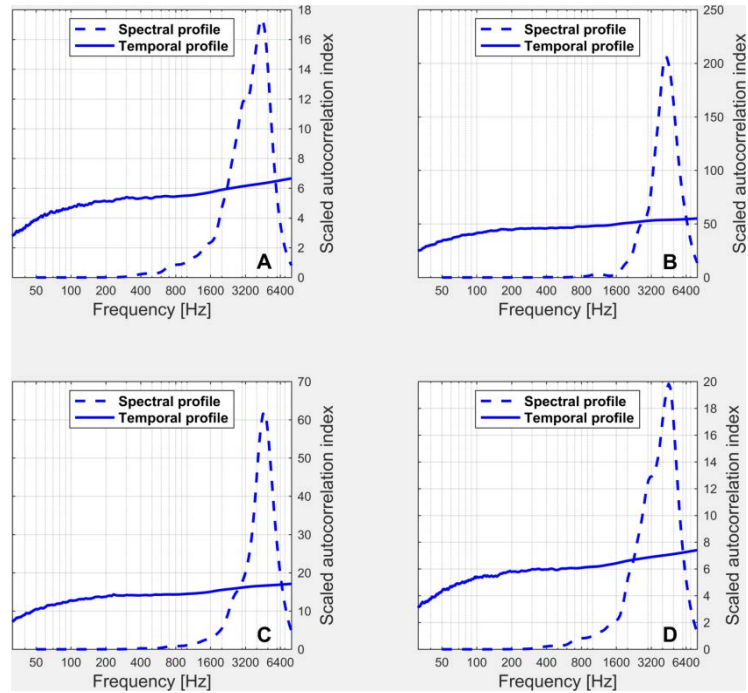

**Fig 16. The dual profile of a 500 ms signal recorded in the conference room** (SN2010 [6]). The solid line is the sum of the ACF along the spectral axis and the dashed line is the sum of the ACF along the time delay axis for 70 ms time interval. The temporal and spectral profiles are scaled to be compared to each other. The 'x' axis is changed in the temporal profile by using the inverse relationship of time to frequency, f=1/t. The sub figure A is for the no object recording and sub figures B, C and D are for the recordings with object at 50, 100 and 200 cm respectively.

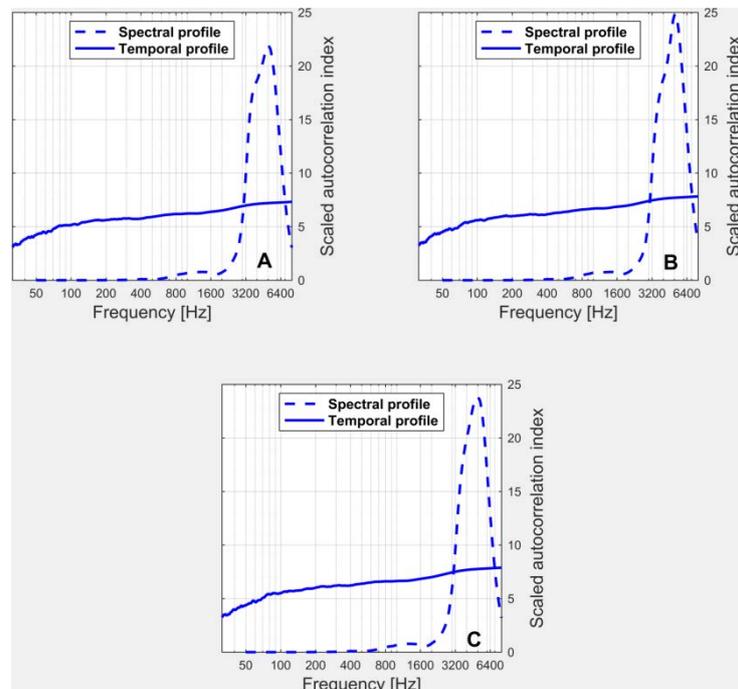

**Fig 17. The dual profile of a 500 ms signal recorded in the lecture room** (SNG2016 [6]). The solid line is the sum of the ACF along the spectral axis and the dashed line is the sum of the ACF along the time delay axis for 70 ms time interval. The temporal and spectral profiles

are scaled to be compared to each other. The 'x' axis is changed in the temporal profile by using the inverse relationship of time to frequency, f=1/t. The sub figure A is for the no object recording and sub figures B and C are for the recordings with object at 100 and 150 cm respectively.

## Pitch strength

The peaks in the temporal profile of the autocorrelation function that we computed with the dual profile module of AIM were distributed without apparent order or meaning. It is not obvious which peak that corresponds to a pitch. There is a solution: The AIM model has a pitch strength module which calculates the pitch strength to determine if a particular peak is random or not. This module first calculates the local maxima and their corresponding local minima. The ratio of peak height to the peak width of the peak (local maxima) is subtracted from the mean of the peak height between two adjacent local minima to obtain the pitch strength (PS) of a particular peak.

Two modifications were made by us in the pitch strength algorithm of AIM to improve its performance for the analysis. (1) The low pass filtering was removed as it smooths out the peaks and, (2) the pitch strength was measured with Equation (14). Figure 18 illustrates the pitch strength algorithm that was used. The peak with the greatest peak height has the greatest pitch strength and would be the perceived frequency of repetition pitch.

$$PS = PH - \overline{PHLM} \qquad (14)$$

where PS is the calculated pitch strength, PH is the height of the peak and $\overline{PHLM}$ is the mean of the peak height between two adjacent local minima.

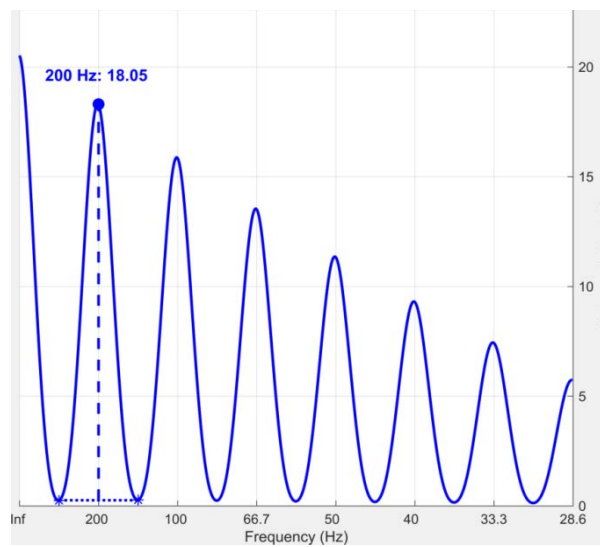

**Fig 18. An illustration of the pitch strength measure computed using the pitch strength module of the AIM.** The dot indicates the local maxima and the two stars are the corresponding local minima. The vertical dashed line is the pitch strength calculated using Equation (14). The frequency in Hz was computed by inverting the time delay, f = 1/t.

The results of the calculated pitch strength for recordings of studies SN2010 [6] and SNG2016 [7] are presented in Tables 6 to 8. As can be seen, peaks were misleadingly identified for recordings without an object, which would not have caused a pitch perception. This happens because the pitch strength algorithm identifies all local maxima and minima in a sound and thus also calculates the pitch strength for all random peaks (that have local maxima).

**Table 6. Mean of the pitch strength (autocorrelation index) of the 10 versions for recordings in SN2010 [6] and in SNG2016 [7] with the 5 ms duration signal.** The blank cells indicate that no recordings were made at those distances.

| | Schenkman and Nilsson [6] | | Schenkman, Nilsson and Grbic [7] |
|---|---|---|---|
| Object distance (cm) | Anechoic room | Conference room | Lecture room |
| No Object, recording 1 | 0.37 | 0.78 | 0.19 |
| No Object, recording 2 | 0.40 | 0.80 | - |
| 50 | 2.54 | 9.65 | - |
| 100 | 1.06 | 2.43 | 0.29 |
| 150 | - | - | 0.28 |
| 200 | 0.35 | 0.85 | - |

**Table 7. Mean of the pitch strength (autocorrelation index) of the 10 versions for recordings in SN2010 [6] with the 50 ms duration signal.**

| | Schenkman and Nilsson [6] | |
|---|---|---|
| Object distance (cm) | Anechoic room | Conference room |
| No Object, recording 1 | 0.47 | 0.55 |
| No Object, recording 2 | 0.44 | 0.52 |
| 50 | 4.17 | 6.59 |
| 100 | 1.67 | 2.22 |
| 150 | 0.42 | 0.54 |

**Table 8. Mean of the pitch strength (autocorrelation index) of the 10 versions for recordings in SN2010 [6] and in SNG2016 [7] with the 500 ms duration signal.** The blank cells indicate that no recordings were made at those distances.

| | Schenkman and Nilsson [6] | | Schenkman, Nilsson and Grbic [7] |
|---|---|---|---|
| Object distance (cm) | Anechoic room | Conference room | Lecture room |
| No Object, recording 1 | 0.71 | 0.84 | 1.30 |
| No Object, recording 2 | 0.78 | 0.90 | - |
| 50 | 4.75 | 7.74 | - |
| 100 | 2.44 | 2.91 | 1.36 |
| 150 | - | - | 1.42 |
| 200 | 0.70 | 1.35 | - |

The unit for pitch strength in our analysis is the autocorrelation index, as it is computed on the autocorrelation function. The tabulated data show that for the 5 ms and 50 ms duration signals the pitch strength was greater than 1 for the object distances of 50 and 100 cm in the anechoic and conference rooms of SN2010 [6] (Tables 6 and 7). For the 500 ms duration signal, the strength was greater than 1 at distances of 50 and 100 cm in the anechoic room and at the distances of 50, 100 and 200 cm in the conference room. The lecture room in SNG2016 [7] also had a pitch strength greater than 1 at this condition, but the computed pitch strength was not consistent over a single frequency and it lasted for only 4 to 8 time frames. (The time frames had 35ms time delay computed from a 70 ms interval NAP signal. Each frame had a hop time of 10ms). This was not the case for the anechoic and conference rooms in SN2010 [6] that both had high pitch strengths at a particular frequency and lasted for 14 to 18 time frames of 35 ms intervals, each with a hop time of 10 ms. Additionally, in the lecture room in SNG2016 [7] with a reflecting object present, the pitch strength was not much different from when there was no object. This further illustrates the echolocating difficulties of the test persons in that study.

The perceptual results of SN2010 [6] showed that the participants were able to detect the objects with a high percentage correct at object distances 50 and 100 cm in the anechoic room and at 50, 100 and 200 cm in the conference room (Schenkman and Nilsson [6]). As presented in the previous paragraph, the pitch strength was greater than 1 at these conditions. Pitch seems to be the most important information that listeners use to detect objects at these distances (see e.g. Schenkman and Nilsson [9]). Therefore, these results imply that there might be a perceptual threshold approximately equal to 1 (autocorrelation index) for pitch strength in echolocating situations. The peak with that pitch strength must exist for certain time frames for a person to perceive a repetition pitch. This is also dependent on the acoustics of the room. Relating the pitch strength results with the performance of the participants in the two studies SN2010 [6] and SN2016 [7] will be made in a following section on threshold values. Before this we shall analyze the results in terms of a timbre property, namely sharpness.

## Sharpness analysis

In the Method part for analysis of room acoustics we described how the spectral centroid was used as a measure for timbre perception. The spectral centroid was computed on the time varying Fourier Transform. To address more specifically how human hearing perceives timbre, Fastl and Zwicker [33] computed the weighted centroid of the specific loudness rather than of the Fourier Transform. This psychophysical measure is called sharpness and indicates how sound extends from being perceived to vary from dull to sharp. We conducted the sharpness analysis for our recordings using code available from Psysound (Cabrera, Ferguson, and Schubert [35]). Sharpness varies over time and therefore the median was used as a representative measure for the perceived sharpness. The results of the mean of the medians of the perceived sharpness over the 10 versions in the anechoic, conference and lecture room in SN2010 [6] and SNG2016 [7] for 5, 50 and 500 ms duration signals are presented in Tables 11 to 14. The unit for sharpness is called acum.

There are, to our knowledge, only a few studies on thresholds of sharpness. Pedrielli, Carletti, and Casazza [46] found that their participants had a just noticeable difference for sharpness of 0.04 acum. You and Jeon [47] found in a study on refrigerator noise that their participants had a just noticeable difference for sharpness of 0.08 acum. Assuming that 0.04 acum is a threshold value for sharpness, then the results in Tables 9 to 11 show that the difference in median sharpness in SN2010 [6] was greater than threshold for the object at 50 and 100 cm compared to the recordings without the object. In SNG2016 [7], the differences between the recordings with and without object were smaller than in SN2010 [6], although greater than 0.04 acum. It is possible that at shorter distances (say less than 200 cm) repetition pitch and loudness information might be more relevant for providing echolocation information than sharpness information.

In study SN2010 [6] with reflecting object at distances 200, 300, 400 and 500 cm for 5 ms (anechoic and conference rooms), 50ms (anechoic and conference rooms) and 500ms signal (conference room) durations, the recordings had differences in mean of median sharpness of less than 0.04 acum when compared to the recordings without an object. However, in the anechoic room of SN2010 for the 500 ms signal duration [6], the recordings with object at 400 cm and 500 cm had differences in sharpness approximately greater than 0.04 acum when compared to the recordings without the object (Tables 9 to 11). This is information that blind persons might use to detect and identify a reflecting object at longer distances than, say at 200 cm. We discuss this issue further in the next section.

**Table 9. Mean of the 10 versions of mean of median of the sharpness (acum) for the recordings in anechoic, conference and the lecture room with the 5 ms duration signal.**

| | Schenkman and Nilsson [6] | | Schenkman, Nilsson and Grbic [7] |
|---|---|---|---|
| Object distance (cm) | Anechoic room | Conference room | Lecture room |
| No Object, recording 1 | 1.89 | 1.97 | 1.85 |
| No Object, recording 2 | 1.9 | 1.98 | |
| 50 | 2.05 | 2.03 | |
| 100 | 2.14 | 2.03 | 1.78 |
| 150 | | | 1.83 |
| 200 | 1.92 | 2 | |
| 300 | 1.91 | 2.01 | |
| 400 | 1.89 | 1.98 | |
| 500 | 1.89 | 1.99 | |

**Table 10. Mean of the 10 versions of mean of median of the sharpness (acum) for the recordings in anechoic, conference and the lecture room with the 50 ms duration signal.**

| | Schenkman and Nilsson [6] | |
|---|---|---|
| Object distance (cm) | Anechoic room | Conference room |
| No Object, recording 1 | 1.89 | 1.89 |
| No Object, recording 2 | 1.9 | 1.89 |
| 50 | 2.07 | 1.96 |
| 100 | 2.14 | 1.95 |
| 150 | | |
| 200 | 1.91 | 1.94 |
| 300 | 1.9 | 1.91 |
| 400 | 1.87 | 1.92 |
| 500 | 1.88 | 1.89 |

**Table 11. Mean of the 10 versions of mean of median of the sharpness (acum) for the recordings in anechoic, conference and the lecture room with the 500 ms duration signal.**

| Object distance (cm) | Schenkman and Nilsson [6] | | Schenkman, Nilsson and Grbic [7] |
|---|---|---|---|
| | Anechoic room | Conference room | Lecture room |
| No Object, recording 1 | 1.86 | 1.94 | 2.07 |
| No Object, recording 2 | 1.88 | 1.94 | |
| 50 | 2.12 | 2.1 | |
| 100 | 2.12 | 2.04 | 2.2 |
| 150 | | | 2.11 |
| 200 | 1.89 | 1.97 | |
| 300 | 1.86 | 1.95 | |
| 400 | 1.83 | 1.95 | |
| 500 | 1.84 | 1.94 | |

# Threshold values, absolute and difference, for echolocation in static situations, based on auditory model analysis

## Background

The stimuli in the studies [6, 7] were presented to the participants in a two-alternative-forced-choice (2AFC) manner. The participants had to compare two sounds and detect the sound with the echo. This perceptual decision is based on information in attributes of the stimuli.

If a person can detect an attribute of an object with a 75 percentage of correct response, this limit is usually termed the absolute threshold. The judgments are commonly based on only one source of information. In the experiments in [6, 7] two sources of information were used. The sounds were presented in a 2AFC manner, where the participants compared the information of two sounds, with and without object, and made the decision. This is a difference threshold (Gescheider [48]), which is the threshold at which the participant can make a discrimination with about 75 percentage of correct response.

As regards echolocation, we define the absolute threshold as the value of a perceptual attribute (e.g. loudness, pitch or sharpness) of a recording with a reflecting object at which a person has a 75 percentage correct response. The difference threshold is thus the difference of responses between the recordings with and without object, respectively, at which the person has 75 percentage of correct response. The experimental procedure used in the experiments was 2AFC, and therefore the difference threshold is the relevant measure of the echolocation ability of the persons. However, as the difference threshold is calculated from absolute thresholds both are for clarity shown here. The procedure for finding the difference thresholds and the corresponding results are presented below.

## Non-parametric versus parametric modeling of psychometric function for threshold analysis

Psychometric functions relate perceptual results to physical parameters of a stimulus. Commonly the psychometric function is estimated by parametric fitting, i.e. it is assumed that the underlying relationship can be described by a specific parametric model. The parameters of such a model are then estimated by maximizing the likelihood. However, the most correct parametric model underlying the description of the psychometric function is unknown.

Therefore, estimating the psychometric function based on the assumptions of a parametric model may lead to incorrect interpretations (Zychaluk and Foster [49]). To address this problem, Zychaluk and Foster [49] implemented a non-parametric model to estimate the psychometric function that does not require assumptions about the empirical state of the underlying phenomenon. The psychometric function is thus modeled locally without assuming a "true" underlying function. Since the true relationship for the variables that determine human echolocation is unknown, we chose the method proposed by Zychaluk and Foster [49] in our analysis of the perceptual data. Next is a brief description of the non-parametric model for estimating the underlying psychometric function. Thereafter our analysis of the perceptual results is presented.

A generalized linear model (GLM) is usually used when fitting a psychometric function. It consists of three components: a random component from the exponential family, a systematic component η and a monotonic differentiable link function g, that relates the two. Hence, a psychometric function, P(x), can be modeled by using Equation (15). The parameters of the GLM are estimated by maximizing the appropriate likelihood function (Zychaluk and Foster [49]). The efficiency of the GLM relies on how much the chosen link function, g , approximates the "true" underlying function.

$$\eta(x) = g[P(x)] \tag{15}$$

However, as mentioned, the true function one can never know for certain. In non-parametric modelling, instead of fitting the link function g, the function η is fitted using a local linear method. For a given point, x, the value η(u) at any point u in a neighborhood of x is approximated by Equation (16) (Zychaluk and Foster [49]).

$$\eta(u) \approx \eta(u) - (u - x)\eta'(x) \tag{16}$$

where $\eta'(x)$ is the first derivative of η. The actual estimate of the value of η(x) is obtained by fitting this approximation to the data over the prescribed neighborhood of x. Two features are important for this purpose, kernel K and the bandwidth h. A Gaussian kernel is preferred, as it has unbounded support and is best for widely spaced levels. An optimal bandwidth can be chosen using plugin, bootstrap or cross validation methods (Zychaluk and Foster [49]). As no method is guaranteed to always work, to find the optimal bandwidth for the analysis we chose a bootstrap method with 30 replications. When the bootstrap method failed to find an optimal bandwidth, cross validation was used to establish the optimal bandwidth.

## Distance thresholds for object detection

The psychometric function was initially fitted to the mean proportion of correct responses with respect to the distance to the reflecting object. Figures 19, 20 and 21 show both the non-parametric modeling (local linear fit) and the parametric modeling of the perceptual results for the blind participants in study SN2010 [6]. The mean percentage correct is plotted as a function of distance for recordings with 5, 50 and 500 ms signals in the anechoic and conference rooms. The link function used for the parametric modeling was the Weibull function. Visual inspection shows that this link function was not appropriate, since the fit does not correspond well with the perceptual results. As mentioned, if one knows the underlying link function for the psychophysical data, then the parametric fit is a better fit than the local non-linear fit, but for the data we are analyzing one cannot assume a particular link function. However, the local linear fit correlates well with the perceptual results. This demonstrates some of the advantages of using non-parametric modeling for the purposes of the present investigation.

We used the means of the proportion of correct responses of the participants for the psychometric fitting. If we had used the individual responses, the individual thresholds would vary but the local linear fit would probably still, we believe, be well correlated with the perceptual results. Therefore, the results in the remaining part of the present section will be based on the psychometric function using local linear fit for the mean proportion correct answers. We used the implementation of the non-parametric model fitting in Matlab by Zychaluk and Foster [49].

The local linear fit needs at least three stimulus values to make a mathematically valid fit. As the recordings in the lecture room (in SNG2016 [7]) had only two stimulus values, at 100 and 150 cm, it was not possible to make a psychometric fit for these recordings.

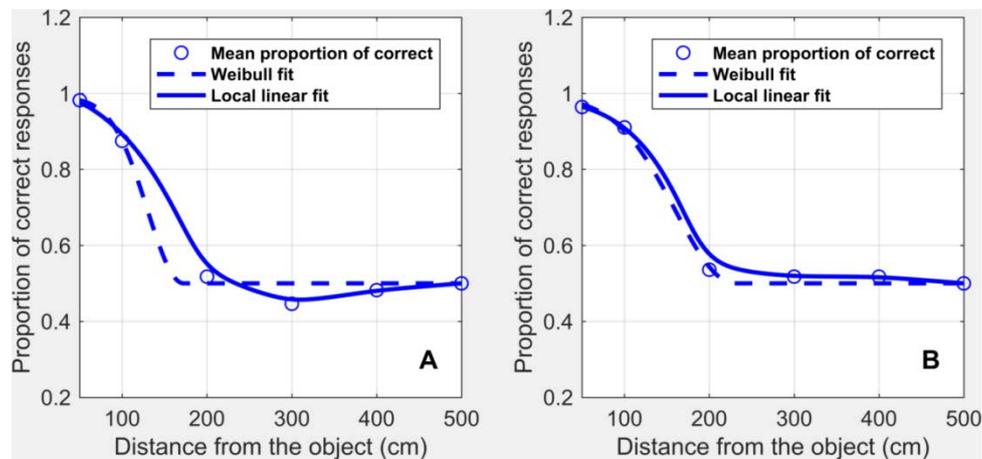

**Fig 19. The parametric (Weibull fit) and non-parametric (Local linear fit) modeling of the mean proportion of correct responses for the 5 ms recordings of the blind participants in SN2010 [6] as a function of distance: (A) anechoic chamber. (B) conference room.**

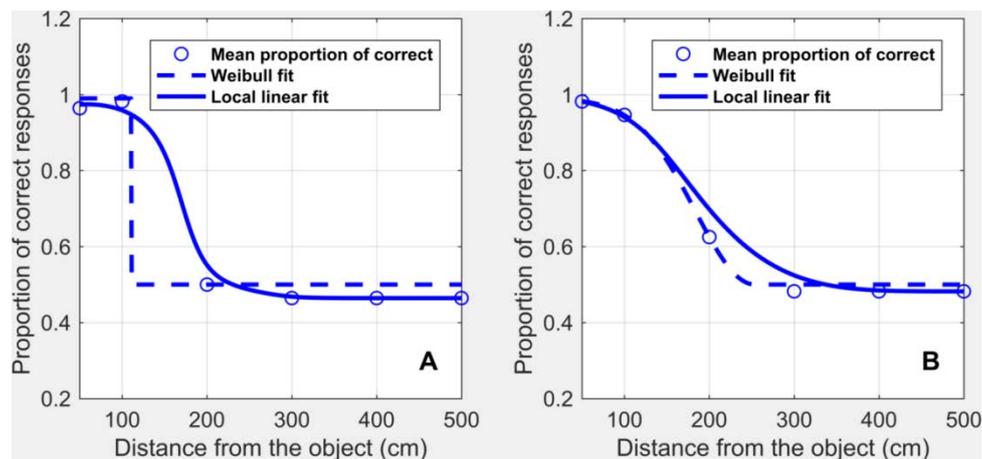

**Fig 20. The parametric (Weibull fit) and non-parametric (Local linear fit) modeling of the mean proportion of correct responses for the 50 ms recordings of the blind participants in SN2010 [6] as a function of distance: (A) anechoic chamber. (B) conference room.**

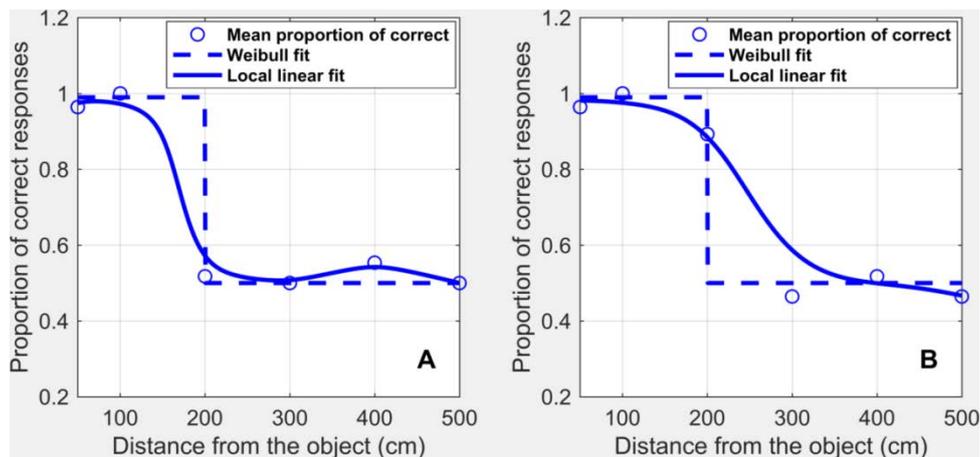

**Fig 21. The parametric (Weibull fit) and non-parametric (Local linear fit) modeling of the mean proportion of correct responses for the 500 ms recordings of the blind participants in SN2010 [6] as a function of distance: (A) anechoic chamber. (B) conference room.**

Distance perception of an object is not a perceptual attribute that was presented directly to the participants in studies SN2010 [6] and SNG2016 [7]. Therefore, the distance threshold obtained from the psychometric fit is a derived quantitative threshold. The distance threshold is the distance at which a person may detect an object with a probability of 75%. As the fitted psychometric function is discrete, it was not always possible for the fit to have an exact value of 0.75. Therefore, the threshold values at the proportion of correct responses within the range of 0.73 to 0.75 were initially calculated and the mean of the threshold values was determined as the actual threshold.

The distance threshold in study SN2010 [6] for which the blind and the sighted could detect the object using echolocation, with a proportion of correct responses between 0.73 and 0.75, are shown in Table 12. The distances at which the blind participants could detect the reflecting object were farther away than for the sighted in both rooms and for all three sound signals. The threshold is positively related to the signal duration for both groups, i.e. the longer durations give a longer range of detection. We can also see that the blind persons could detect objects farther away in the conference room for all signals, but for the sighted this was only the case for the 500 ms signal. In the original study SN2010 [6, table 3], the calculations of at what distance a blind or a sighted person might detect a reflecting object were based on a parametric approach, yielding in general lower distance values for thresholds than our non-parametric approach. In other words, our reanalysis of the data show a higher sensitivity of both blind and sighted, than the values in [6].

**Table 12. Distance thresholds (cm) for duration, room, and listener groups in study SN2010 [6].** The threshold values were calculated from the psychometric function of the blind and sighted participants' responses at the mean percentage of correct responses values of 0.73 to 0.75.

| | Sound duration (ms) | | | | | |
|---|---|---|---|---|---|---|
| | 5 | | 50 | | 500 | |
| Room | Blind | Sighted | Blind | Sighted | Blind | Sighted |
| Anechoic | 150 | 130 | 166 | 160 | 172 | 166 |
| Conference | 158 | 121 | 176 | 147 | 247 | 207 |

## Loudness thresholds, absolute and difference, for object detection

Loudness is also a source of information for detecting reflecting objects by echolocation (Schenkman and Nilsson [9]; Kolarik et al, 2014 [2]). A common psycho-acoustical measure to express loudness is in the unit of sones (Yost [50]; Gescheider [48]). Therefore, we used values in sones for the local linear fitting to determine the absolute and difference threshold of loudness, where blind and sighted could detect a reflecting object. As elsewhere, the criterion was detection with a percentage correct between 73 and 75%. The mean loudness values, and the mean percentage of correct responses calculated from study SN2010 [6] were used as inputs for the psychometric fit. The resulting absolute threshold values of loudness for detecting the object are presented in Table 13. The data show that the thresholds for loudness for the blind participants were lower compared to those of the sighted, roughly 1 sone less in the anechoic chamber and 2 sones less in the conference room.

**Table 13. Absolute threshold values of loudness (sones) for duration, room, and listener groups in study SN2010 [6].** The threshold values were calculated from the psychometric function of the blind and sighted participants' responses at the mean percentage of correct responses value of 0.73 to 0.75.

| | Sound duration (ms) | | | | | |
| | 5 | | 50 | | 500 | |
| Room | Blind | Sighted | Blind | Sighted | Blind | Sighted |
|---|---|---|---|---|---|---|
| Anechoic | 16.8 | 17.5 | 43.7 | 45.1 | 52.9 | 53.2 |
| Conference | 22.6 | 24.1 | 49.4 | 53.1 | 53.6 | 55.3 |

The values in Table 13 may misleadingly lead a reader to infer that the shortest sounds had the lowest threshold. This is not the case. If we look at tables 3 and 5 the recording without object for the 5 ms signal had a loudness of approximately 13 sones and the 500 ms signal had approximately 45 sones. Considering these values, it is more appropriate to use the difference rather than the absolute threshold, since the detection based on loudness in SN2010 [6], as a consequence of the 2AFC method that had been used, was based on a relative judgment, a comparison, and not on an absolute judgment. The difference threshold values were calculated by subtracting the absolute threshold values in Table 13 with their corresponding loudness values for the recording without object, see Table 14.

**Table 14. Difference threshold values of loudness (sones) for duration, room, and listener groups in study SN2010 [6].** The threshold values were calculated from the psychometric function of the blind and sighted participants' responses at the mean percentage of correct responses value of 0.73 to 0.75.

| | Sound duration (ms) | | | | | |
| | 5 | | 50 | | 500 | |
| Room | Blind | Sighted | Blind | Sighted | Blind | Sighted |
|---|---|---|---|---|---|---|
| Anechoic | 3.4 | 4.1 | 3.6 | 5 | 4.8 | 5.1 |
| Conference | 3.2 | 4.7 | 4.3 | 8 | 1.1 | 2.8 |

The data in Table 14 give a different and more realistic picture than the data in Table 13. The blind detect objects at lower loudness values for all sounds, and both groups could detect with lower relative loudness levels (loudness difference between the recording with and without object) for 500ms duration signals in the conference room. The loudness model used to compute the mean loudness was the same for both test groups, and therefore we conclude that the low thresholds of the blind persons are an effect of their perceptual ability. This conclusion is further discussed in the Discussion part of this report.

## Pitch thresholds, absolute and difference, for object detection

The absolute and difference threshold values of pitch strength, as calculated by the autocorrelation index for which the blind and the sighted test persons in study SN2010 could detect the reflecting object, are presented in Tables 15 and 16, respectively. We will first discuss the results in terms of the absolute thresholds, but the more appropriate conclusions will be based on the difference thresholds.

The absolute threshold varies for blind and sighted persons depending on signal durations and room conditions. The blind, for all conditions, had lower thresholds, the pitch strength increased with signal duration and the thresholds were lower in the anechoic room. It is possible that for shorter duration signals, the person may be inattentive and miss the signal and thus also the pitch information. The performance (percentage of correct response) of the participants with 5 and 50 ms signals may thus not only be based on pitch strength but also on cognitive factors such as attention.

Schenkman and Nilsson [9] showed that when pitch and loudness information were presented together, at distances up to 200 cm to the object, the participants' performance was almost 100 percentage correct. The 500 ms recordings with the object at 50 and 100 cm in study SN2010 [6] had almost 100 percent correct response for both the blind and the sighted. Therefore, for the 500 ms signal condition it is likely that the likelihood to miss a signal and its pitch information because of non-attention is lower, and the perceptual results of the participants are likely to be based mostly on pitch information.

There are two possible theoretical ways to regard how the hearing system treats the ACF values. We will here focus the analysis on the 500 ms signal, since as noted above, the 5 ms and 50 ms signals may have cognitive aspects that could bias the auditory model analysis. (1) Based on the above reasoning, and if we assume that the auditory system analyses the pitch information absolutely i.e. it does not compare the peak heights in the ACF between the recordings (when presented in a 2AFC manner), then the results indicate that the absolute threshold for detecting a pitch based on an autocorrelation process should be greater than 1.10 and 1.23 (as indicated by the autocorrelation index for the 500 ms signal) for the blind and the sighted, respectively, as shown in Table 15. (2) On the other hand, if we assume that the auditory system analyses the pitch information relatively i.e. it compares the peak heights in the ACF between the recordings (when presented in a 2AFC manner) then the results indicate that the difference threshold for detecting the pitch based on autocorrelation should be greater than 0.27 and 0.49 (autocorrelation index) for the blind and the sighted, respectively, as shown in Table 16. For all cases, the values show that the blind persons could detect echo reflections of objects having lower peak heights in the ACF, than the sighted could.

**Table 15. Absolute threshold values of pitch strength (autocorrelation index) for duration, room, and listener groups in study SN2010 [6].** The threshold values were calculated from the psychometric function of the blind and sighted participants' responses at the mean percentage of correct response values of 0.73 to 0.75.

| | Sound duration (ms) | | | | | |
|---|---|---|---|---|---|---|
| | 5 | | 50 | | 500 | |
| Room | Blind | Sighted | Blind | Sighted | Blind | Sighted |
| Anechoic | 0.77 | 0.88 | 0.8 | 0.96 | 1.1 | 1.23 |
| Conference | 1.54 | 2.21 | 1.07 | 1.69 | 1.14 | 1.41 |

**Table 16. Difference threshold values of pitch strength (autocorrelation index) for duration, room, and listener groups in study SN2010.** The threshold values were calculated from the psychometric function of the blind and sighted participants' responses at the mean percentage of correct response values of 0.73 to 0.75.

| | Sound duration (ms) | | | | | |
|---|---|---|---|---|---|---|
| | 5 | | 50 | | 500 | |
| Room | Blind | Sighted | Blind | Sighted | Blind | Sighted |
| Anechoic | 0.39 | 0.50 | 0.35 | 0.51 | 0.36 | 0.49 |
| Conference | 0.75 | 1.42 | 0.54 | 1.16 | 0.27 | 0.54 |

## Sharpness thresholds, absolute and difference, for object detection

As discussed previously, timbre indicates how we experience various qualities of sounds. We chose to study one aspect of timbre, sharpness, as a potential information source for object detection by echolocation. In analog to the previous psycho-acoustical parameters, we calculated the absolute and difference threshold values of sharpness for which the blind and the sighted test persons in study SN2010 [6] could detect a reflecting object using echolocation with a correct response value of 0.73 to 0.75.

For quantitative values for sharpness, we used the psychophysical unit acum (e.g. Fastl and Zwicker [33]). Tables 17 and 18 show that for the blind and sighted participants both their absolute and difference thresholds, the sharpness values were about the same. However, unlike loudness and pitch strength the sharpness information need not be greater in value for the participants to detect the object. For sharpness, a listener must distinguish between timbres. This may include cognition, involving e.g. memory processes.

When a participant in SN2010 [6] and SNG2016 [7] were presented with two stimuli in a 2AFC method, they distinguished the recording with the object from the recording without the object by identifying the one with the higher loudness level, stronger pitch strength or both. However, when a person uses sharpness for echolocation, e.g. in a 2AFC method, it is not necessary that the recording with the reflecting object has the higher sharpness value. The recording with the reflecting object might sound duller, i.e. having a lower value of sharpness, than the recording without the object. A person might use this information to detect or identify an object.

**Table 17. Absolute threshold values of the mean of the median sharpness (in the unit acum) for duration, room, and listener groups in study SN2010 [6].** The threshold values were calculated from the psychometric function of the blind and sighted participants' responses at the mean percentage of correct response values of 0.73 to 0.75.

| | Sound duration(ms) | | | | | |
|---|---|---|---|---|---|---|
| | 5 | | 50 | | 500 | |
| Room | Blind | Sighted | Blind | Sighted | Blind | Sighted |
| Anechoic | 1.97 | 1.98 | 1.96 | 1.98 | 1.94 | 1.96 |
| Conference | 2.01 | 2.03 | 1.94 | 1.94 | 1.97 | 1.97 |

**Table 18. Difference threshold values of the mean of the median sharpness (in the unit acum) for duration, room, and listener groups in study SN2010 [6].** The threshold values were calculated from the psychometric function of the blind and sighted participants' responses at the mean percentage of correct response values of 0.73 to 0.75.

| | Sound duration(ms) | | |
|---|---|---|---|
| | 5 | 50 | 500 |

| Room | Blind | Sighted | Blind | Sighted | Blind | Sighted |
|------|-------|---------|-------|---------|-------|---------|
| Anechoic | 0.07 | 0.08 | 0.06 | 0.08 | 0.07 | 0.09 |
| Conference | 0.03 | 0.05 | 0.05 | 0.05 | 0.03 | 0.03 |

Table 11 showed that the sharpness values for the 500 ms duration recordings, with reflecting object at 400 and 500 cm in the anechoic room, had smaller sharpness values than the recording without an object. Interestingly, two blind participants (no. 2 and 6) performed better at these conditions than all the remaining participants, i.e. their proportion correct were approximately 0.7, even at 400 and 500 cm. We looked deeper into the performance of these two high-performing echolocators by making a local linear fit for the proportion correct of these two persons and the sharpness values of the 500ms recordings in the anechoic chamber that were calculated and are shown in Table 11. Cross validation was used to find the bandwidth of the local linear fit kernel. Figs 22 and 23 show the corresponding local linear fits.

One can see clearly that when the proportion correct was approximately equal to 0.7, there were two absolute threshold values for sharpness, one higher and one lower. If we consider the mean of the sharpness of the two no object recordings at this condition (i.e. anechoic room and 500ms signal duration) it was about 1.87 (Table 11). Hence the difference threshold for the blind participant 2 would be 1.94 -1.87 = 0.07 acum and 1.83-1.87 = -0.04 acum. Similarly, the difference threshold for the blind participant 6 would be 1.97-1.87 = 0.10 acum and 1.83-1.87 = -0.04 acum. Perceptually, this means that the two high-performing blind participants could detect the object even when the recording with the object was duller than the recording without an object. A more detailed discussion on the importance of sharpness information for human echolocation is presented in the Discussion part below.

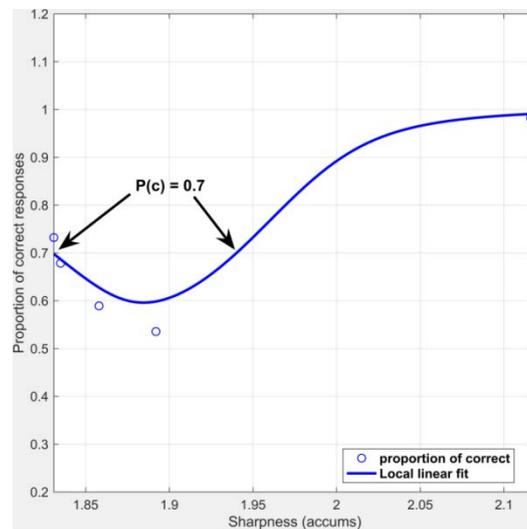

**Fig 22. The non-parametric (local linear fit) modeling of the proportion of correct response for the 500 ms recordings of the 2nd blind participant in SN2010 [6] as a function of the mean of median sharpness.** As the function is discrete, the values of proportion correct responses approximately equal to 0.7 are considered.

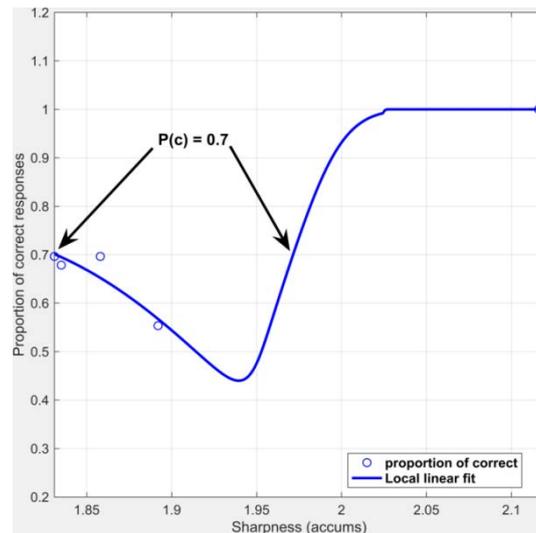

**Fig 23. The non-parametric (local linear fit) modeling of the proportion of correct response for the 500 ms recordings of the 6<sup>th</sup> blind participant in SN2010 [6] as a function of the mean of median sharpness.** As the function is discrete, the values of proportion correct responses approximately equal to 0.7 are considered.

# Discussion

We wanted to elucidate factors that determine echolocation and the differences of echolocation ability between blind and sighted persons. Neuroimaging and psychoacoustic methods give us insight into the high echolocating ability of blind persons, but these methods do not necessarily reveal the information in the acoustic stimulus that determines echolocation (at least when the information is not known) and also not fully how this information is represented and processed in the human auditory system. The implementation of auditory models for human echolocation was intended to establish the information that determines echolocation ability and its variability, and how this information is represented and processed in the human auditory system.

Signal analysis was conducted on the physical signals and is presented in the Method part on Room acoustics. The aim was to find the physical information that is used for echolocation and to analyze the effects of room acoustics on human echolocation. We studied in particular the sound pressure levels, auto correlations and spectral centroids. The analyses were performed on the recordings of studies SN2010 [6] and SNG2016 [7]. The results demonstrate, as expected, that the acoustics of the room affect the sounds and thereby the physical attributes that are associated with it. However, the information represented in the auditory system is complex and it uses this physical information for further processing through the auditory neural system. To understand better what takes place in the auditory system of a person using echolocation, we used what we consider are the most relevant auditory models for human echolocation, that today are available in the literature. We thus studied how the corresponding perceptual attributes of sound pressure level, auto correlation and spectral centroid are encoded in the human auditory system.

The results of the auditory models suggest that repetition pitch, loudness and sharpness all provide potential information for people to echolocate at distances below 200 cm. The results also indicate that at longer distances sharpness information may be used for human echolocation. A detailed discussion of how loudness, pitch and sharpness are essential for human echolocation and how they might be represented in the auditory system is presented below in sections "Echolocation and loudness", "Echolocation and pitch" and "Echolocation and timbre". A discussion of how room acoustics and binaural information as well as movement affect human echolocation is presented in the section "Echolocation and room acoustics" and "Echolocation, binaural information and movement". We finally conclude by some comments on using auditory models for understanding human echolocation together with some theoretical implications.

## Echolocation and loudness

Of the existing loudness models, we chose the one by Glasberg and Moore [31], since it has a good fit to the equal loudness contours in ISO 2006. The results of the model were related to the proportion of correct responses of the listeners in studies SN2010 [6] and SNG2016 [7] for calculating estimates of threshold values based on loudness. The loudness values are tabulated in Tables 3 to 5 and the resulting threshold values for detecting a reflecting object, when based on the percentage correct, are shown in Tables 13 and 14.

The difference in loudness level between the loudness threshold and loudness level of the recordings without the object for 5, 50 and 500 ms duration signals in the anechoic room were approximately 4.2, 5 and 5 sones for the sighted persons. As an example, the 5 ms signal in the anechoic room Table 13 had a threshold of 17.5 sones for the sighted persons, while the mean for loudness with no object in this room was 13.3 sones, as shown in Table 3. The difference of these two values, 17.5 and 13.3, is 4.2 sones. For the conference room those differences were 5, 8 and 3 sones, respectively (Tables 3 to 5 and Table 13). These differences in loudness level make it possible for persons to echolocate, making loudness a potential information source for echolocation (see also Schenkman and Nilsson [9]; Kolarik et al, [2]). Comparing the loudness thresholds of sighted and blind persons, the thresholds of the blind persons were lower than those of the sighted test persons (Table 13). If loudness information is encoded in the same manner for both groups of test persons, which we believe is a reasonable assumption, then this analysis shows that blind persons may echolocate at lower loudness levels than sighted persons.

## Echolocation and pitch

Repetition pitch is one of the important information sources that blind people use to detect a reflecting object at shorter distances (e.g. Bilsen [14]; [6]). We studied how this information is represented in the auditory system. For this purpose, a dual profile analysis was performed and is presented above in the section "Pitch analysis: autocorrelation with dual profiles". The results suggest that repetition pitch can be explained by the peaks in the temporal profile rather than by peaks in the spectral profile of the autocorrelation function. This is in agreement with a study by Yost [26], where the peaks in the temporal domain of the autocorrelation function form the basis for explaining the perception of repetition pitch.

However, the dual profile analysis was not sufficient to determine the strength of the perceived pitch as the peaks were more random in the temporal profile of the autocorrelation function. A measure of pitch strength was therefore used that showed if the peaks were random or not and thereafter computed the pitch strength (Equation 14). The means of the resulting pitch strengths are shown in Tables 6 to 8 and the thresholds for detecting objects in Tables 15 and 16. Only the pitch strength values obtained from the 500 ms duration recordings in Table 8 were considered, as these recordings are not likely to be influenced by cognitive factors. The pitch strength threshold was lower for blind persons than for sighted. A reasonable assumption is that pitch information is encoded in the same manner for both blind and sighted persons. Then it appears that blind persons may echolocate with a lower pitch strength than sighted persons. The auditory models were used without changing its parameters for the analysis of the two groups and it is thus not possible to infer what factors that determine the perceptual differences.

## Echolocation and timbre

To determine to what extent that sharpness is useful for echolocation, we computed the weighted centroid of the sounds from studies SN2010 [6] and SNG2016 [7] for specific loudness using the code of Psysound3. Pedrielli, Carletti, and Casazza [46] showed in their analysis that the just noticeable difference for sharpness in their study was 0.04 acum. We used this value as a criterion of sharpness for detecting reflecting objects. The results from our analysis (Tables 9 to 11) show that the difference in sharpness was greater than 0.04 acum for recordings with the object at distances of 50, 100, 150 and 200 cm. However, at these distances both loudness and pitch information are more prominent. Hence, at distances shorter than 200 cm, sharpness might not be a major information source for echolocation.

One may note that, in study SN2010 [6], for the 500 ms recording in the anechoic chamber, with the reflecting object at 400 cm and 500 cm, the sharpness difference was approximately equal to 0.04 acum when compared to the mean of the two recordings without the object (Table 11). A few of the blind test persons in SN2010 were able to detect objects at 400 cm. If the just noticeable difference for sharpness of 0.04 acum as found by Pedrielli, Carletti, and Casazza [46] also is a difference threshold for sharpness when echolocating sounds at longer distances than about 2 m, then sharpness can be used as vital information for blind people to detect objects at 400 cm. Here we point out that Tables 17 and 18 present a linear relationship between sharpness and percentage correct, i.e. if there is a higher value of sharpness then a higher probability of detection will result. However, as discussed previously, in distinction to loudness or pitch, sharpness does not need to be larger to indicate detection, since this may be indicated by either being perceived as dull or sharp. A psychometric function cannot depict this. An experiment controlling for sharpness information of the sound could clarify its role for echolocation.

## Echolocation and room acoustics

Loudness, pitch and sharpness provide people with information useful for echolocation, but the efficiency of these also depend on the acoustics of the room and the character of the sounds. The results of many studies of human echolocation are evidence of this (Kolarik et al [2]; Tonelli, Brayda and Gori [52]; Vercillo et al [53]).

The conference room in SN2010 [6] increased pitch strength and hence enabled the participants to echolocate at farther distances, while the lecture room in SNG2016 [7] decreased pitch strength and the listeners presumably had to rely more on other kinds of information like loudness, resulting in deterioration of object detection. The physical cause for the deterioration was probably because in SN2010 [6] the loudspeaker was on the chest of the artificial head, while in SNG2016 [7] it was 1 m behind the artificial head. Another physical cause for the deterioration might be that the reverberation time in the conference room in SN2010 [6] was 0.4s, but in the lecture room of SNG2016 [7] it was 0.6s. Too little reverberation does not seem to be beneficial for human echolocation (see also Tonelli, Brayda and Gori [52]), but too much cannot neither be useful. We hypothesize that there might be an optimal amount of reverberation for successful echolocation. Careful design of room acoustics should improve the possibilities for echolocation by blind persons.

The effects of room acoustics for echolocation are also provided by the recordings in the anechoic room in study SN2010 [6]. The recordings with object at all distances in this room, including those at 400cm and 500cm, had no other reflections than from the reflecting object. The slight sharpness differences that resulted might be used by very skilled listeners to detect an object.

## Echolocation, binaural information and movement

Binaural information may provide additional information for echolocation. Both inter-aural level differences and time differences provide information for echolocation. Papadopoulos et al. [54] argued that information for obstacle discrimination were found for the frequency dependent inter aural level differences especially in the range from 5.5 to 6.5 kHz. Nilsson and Schenkman [55] found that the blind people in their study used interaural level differences more efficiently than the sighted.

There is evidence that self-generated sounds (e.g. Kellogg [56]; Rice, [57]), as well as binaural information (e.g. Dunai, Peris-Fajarnés and Brusola [58]) is beneficial for echolocation. The recordings of the analyzed studies in this report, SN2010 [6] and SNG2016 [7], had the reflecting object directly in front of the recording microphones of the dummy head, and very little binaural information was therefore provided to the test persons. It was thus not considered as a source of information in this report, although we did calculate the SPL values at each ear. As can be seen in Tables 1 and 2, the SPL values at both ears were very similar.

We may here add that the static nature of the recordings might have resulted in a poorer echolocation of the test persons. In a real situation movement gives additional information. Blind persons may move their heads and body, or the object might be moving. Additionally, they would also use their own sounds. It is reasonable to conclude that such sounds, together with motion, offer more information to blind persons than static sounds that are provided by an external source or by an experimenter [2-3].

The present article is based on two studies, both using stationary objects and stationary listeners. Already Wilson [4] showed theoretical reasons for the benefits for echolocation that could be achieved through motion. The same conclusions are obtained by the findings of Bassett and Eastmond [22] of how pitch changes depending on the distance to the reflecting object. More recent findings by Rosenblum et al [59] showed advantages for walking when echolocating. Furthermore, self-motion has been found to be beneficial for echolocation (Wallmeier and Wiegrebe [5]). In a study on blind children walking along a path, Ashmead, Hill and Talor [60] found that the children could avoid a box by utilizing non-visual information. These authors stated that congenitally blind children could utilize some of the auditory information and that they could coordinate this information with functionally important behavior, such as goal-directed locomotion.

Arias et al [61] saw echolocation as a combination of action and perception, while Thaler and Goodale [3] in their review stressed that echolocation is an active process. Our present analysis should be extended to also include studies of motion of objects or persons, as well as of self-generated sounds. The same kinds of auditory models, possibly with some modification, could then be used to test processes or hypothesis pertaining to the additional benefits of motion and of self-generated sounds for echolocation.

## Comments on the auditory model approach to human echolocation

In psychoacoustic experiments a sound is usually presented to participants in a controlled manner and the perceptual or behavioral responses are measured. This makes it possible to identify cause-effect relationships between stimuli and response. However, although the stimuli are presented in a controlled manner in e.g. echolocation studies, the underlying cause for the echolocation is not obvious, e.g. whether it is biological, perceptual or psychological. For example, in the study SN2010 [6], the blind participants were able to perform better than the sighted, but the main cause for the high performance is not evident, except that the cause is related to blindness.

De Volder et al [62] were maybe the first to describe different brain activities in blind and sighted persons in distance tasks by using positron emission tomography. Other advanced scanning techniques like functional magnetic resonance imaging of the brain may locate areas in the brain that are activated when persons use echolocation. These techniques can describe physiological processes relating to or underlying echolocation abilities (Thaler et al [63]; Thaler et al [64]), but such analyses do not fully reveal how the information used by blind persons for echolocation is processed and represented in the auditory system, how it cognitively is perceived and analyzed. Recently there has also been some serious criticism against MRI studies, that their results may be inflated by false-positive rates (Eklund, Nichols and Knutsson [65]).

To address the issues of representation and processing, we implemented a number of auditory models; the binaural loudness model of Moore and Glasberg [32], the auditory image model, AIM, of Patterson, Allerhand, and Giguere [21] and the sharpness model of Fastl and Zwicker [33]. We chose the loudness model of Moore and Glasberg [32] since it agrees well with the equal loudness contours of ISO 2006 and also gives an accurate representation of binaural loudness (Moore [44]). We chose the auditory image model, AIM, since instead of using two different modules to depict frequency selectivity and compression; it uses a dynamic compressive gammachirp filterbank (dcGC) module to depict both the frequency selectivity and the compression performed by the basilar membrane. The AIM model of Bleeck, Ives, and Patterson [34], was thus used to analyze the information provided by repetition pitch. This model is also physiologically inspired. Finally, we used the loudness model of Glasberg and Moore [31] to analyze sharpness. The sharpness information was obtained from the weighted centroid of the specific loudness (Fastl and Zwicker [33]). A general review of auditory processing models is given by Dau [8].

The signal analysis performed on the physical stimuli showed how sound pressure level, autocorrelation and spectral centroid varied with the recordings. The results with AIM showed that the peaks in the temporal information was the likely source for echolocation at shorter distances, and this explanation is thus in line with the analysis by Bilsen [14] and Yost [26] in how the perception of repetition pitch is represented in people, i.e. that the information necessary for pitch perception is represented temporally in the auditory system. The analysis performed with the sharpness model showed that blind participants in our analyzed studies could have used sharpness to detect objects at longer distances and that both temporal and spectral information are required to encode this attribute. Our analysis has some similarities to that of Rowan et al [66] in utilizing models to analyze the perception of level information. Similar to their analysis we saw cross channel cues or spectral spread information as relevant for object detection, which we used in our quantification of sharpness.

The analyses with the auditory models we have used do not fully explain how information necessary for the high echolocation ability of blind persons is represented in the auditory system. In order to investigate by auditory models, the neural and physiological causes to the different echolocation abilities of blind and sighted people, the parameters of the models could be varied to fit participants´ perceptual results. However, we have assumed that the high echolocation ability was due mainly to a perceptual ability common to both groups. Therefore, the thresholds for the blind and the sighted persons were obtained by comparing the results of the auditory models with the perceptual results of both test groups in SN2010 [6] and SNG2016 [7]. As mentioned, the blind participants were consistently better than sighted when echolocating, and they had lower thresholds of detection than the sighted for all parameters that were studied.

The analysis with the auditory models confirmed that repetition pitch and loudness are important information sources for people when echolocating at shorter distances, which is in agreement with earlier results (e.g. Schenkman and Nilsson [6, 9]; Kolarik, Cirstea, Pardhan, and Moore [2]). Sharpness is a candidate for being an important source for echolocation both at short and long distance. Psychoacoustic experiments could determine the usefulness of sharpness and other timbre qualities for echolocation. Highest ecological validity is probably reached by experiments with real objects in actual environments, but laboratory studies are a viable alternative. Today simulations of rooms and objects provide another option for how to study human echolocation, see e.g. Pelegrin-Garcia, Rychtáriková and Glorieux [66].

## Conclusions

We used auditory models to analyze how information for human echolocation in static situations is represented and processed in the auditory system, i.e. when no movement is involved. We focused on three perceptual attributes. Two of these, loudness and pitch, are known to be important for human echolocation. The third attribute, sharpness, an aspect of timbre, we considered important for echolocation and was therefore also studied. We used a number of auditory models: The binaural loudness model of Moore and Glasberg [32], the AIM model of Bleeck, Ives, and Patterson [34], the loudness model of Glasberg and Moore [31], and for sharpness information we used the weighted centroid of the specific loudness, as formulated by Fastl and Zwicker [33].

The main results of our analysis are the following. At shorter distances between person and reflecting object, repetition pitch, loudness and also sharpness provide information to detect objects by echolocation. At longer distances, sharpness information might be used for the same purpose. This tentative conclusion has to be justified experimentally by varying in particular the sharpness characteristics of echolocation sounds. Our analysis confirmed that repetition pitch can best be represented in the auditory system by the peaks in the temporal profile rather than by the spectral profile (see also Yost [26]). As the median sharpness information is computed by using the centroid of the specific loudness varying over time, it is represented by both the spectral and temporal information.

For our analysis we have assumed that the auditory information for both blind and sighted persons is represented and processed in the same way. However, this assumption may not be true. The high echolocation ability of the blind may be the outcome of physiological changes in the neural system, as some studies have indicated (Thaler et al [63]; Thaler et al [64]). To investigate this in further detail, one should change the parameters of the auditory models and analyze the results together with data from neuroimaging and psychoacoustic experiments. If it is established that the underlying ability of blind persons is because of certain physiological conditions, then the parameters of the auditory models can be varied until the results from the auditory models agree with the psychoacoustic results.

We used different auditory models to analyze loudness, pitch and sharpness attributes. It was assumed that the high echolocation ability of many blind persons is a result of a general perceptual ability, and we therefore computed perceptual thresholds for the blind and the sighted persons by using the same models for both groups. The analysis showed that the blind had lower thresholds than the sighted and could echolocate at both lower loudness and lower pitch strength levels. As noted above, the recordings in SN2010 [6] and SNG2016 [7], that form the basis for our analysis were recorded in static positions, i.e. there was no movement between the person and the object. In real life, a blind person would likely be moving, or the object would be so. In addition, the person is often using his/her own sounds, which is advantageous for echolocation (e.g. Kellogg, [56]; Rice, [57]). When movement is provided, and self-produced sounds are used, we believe that the thresholds for the blind persons would be even lower. These ideas are in alignment with the concept of surplus information in Schenkman and Nilsson [6] that more information makes perceptual tasks easier to perform, while lack of information makes perception ambiguous and difficult. This concept follows from Gibsons´s [67] theory of ecological perception. In summary, we have shown the importance of pitch, loudness and timbre for human echolocation. These three characteristics have to be further studied, but especially the role of timbre attributes like sharpness, needs a deeper understanding. Neuroimaging, psychoacoustic experiments, and signal analysis including auditory models, may help us to understand how information necessary for the high ability of many blind persons or with visual impairments is represented and perceived.

## Acknowledgment

The original data were collected together with Mats E. Nilsson. This work was partially supported by the Swedish Council for Working Life and Social Research, (grant number 2008-0600) https://forte.se/, and by Promobilia (grant number 12006) https://www.promobilia.se/, both to BS. The funders had no role in study design, data collection and analysis, decision to publish, or preparation of the manuscript.

## Author contributions

Conceived and controlled the study: BS VG. Performed the simulations: BS VG. Analysed the data: BS VG. Wrote the paper: BS VG. Funding acquisition: BS. Contributed analysis tools: VG. Developed the theory: BS VG. Wrote the paper: BS VG.

**S1 Supporting information: Calculation of calibration constant**

The reference sound pressure level (SPL) for the calibration constants in the anechoic, conference and lecture rooms in SN2010 [6] and SNG2016 [7], were documented in dB(A), 77, 79 and 79 dB(A), respectively. The calibration constant of the recordings should, in principle, be A weighted. The results in Table S.1 show the calibrated levels calculated using the calibration constants in Equations A.1 and A.2, respectively. The A weighted signal gives an increase in calibrated levels by approximately less than 0.5 dB, which is negligible for human hearing. We therefore used Equation A.1, and not Equation A.2, for calculating the calibration constants for all recordings in this report.

$$CC = 10^{\left(\frac{SPL - 20 * log10\left(\frac{rms(signal)}{20 * 10^{-6}}\right)}{20}\right)}$$

(A.1)

$$CC = 10^{\left(\frac{SPL - 20 * log10\left(\frac{rms(Aweighting(signal))}{20 * 10^{-6}}\right)}{20}\right)}$$

(A.2)

The results of Equations A.1 and A.2 for the calibrated levels with and without A weighting were calculated for the recordings of the 9th version of the left ear, 500 ms signal with no object of the first recording in the anechoic and conference rooms and of the 9th version of the left ear, 500 ms with no object recording in the lecture room, see Table S.1.

|  | Schenkman and Nilsson [6] | | Schenkman, Nilsson and Grbic [7] |
|---|---|---|---|
|  | Anechoic room | Conference room | Lecture room |
| Without A weighting: Equation A1 | 77.0 | 79.0 | 79.0 |
| With A weighting: Equation A2 | 77.5 | 79.5 | 79.3 |

**Table S.1:** Calibrated levels with and without A weighting for the 9th version at the left ear for a 500ms sound with no reflecting object, first recording, in anechoic and conference rooms, and for the 9th version of left ear for 500ms sound with no reflecting object in lecture room.

**S2 Supporting information: Pitch strength using strobe temporal integration**

The temporal profile of the stabilized auditory image for a recording of a 500 ms in the conference room in SN2010 [6] is shown in Fig S.2. As stated in the section "Auditory analysis of acoustic information", the stabilized auditory image was implemented with two modules, sf2003 and ti2003. A brief description of this implementation is given below.

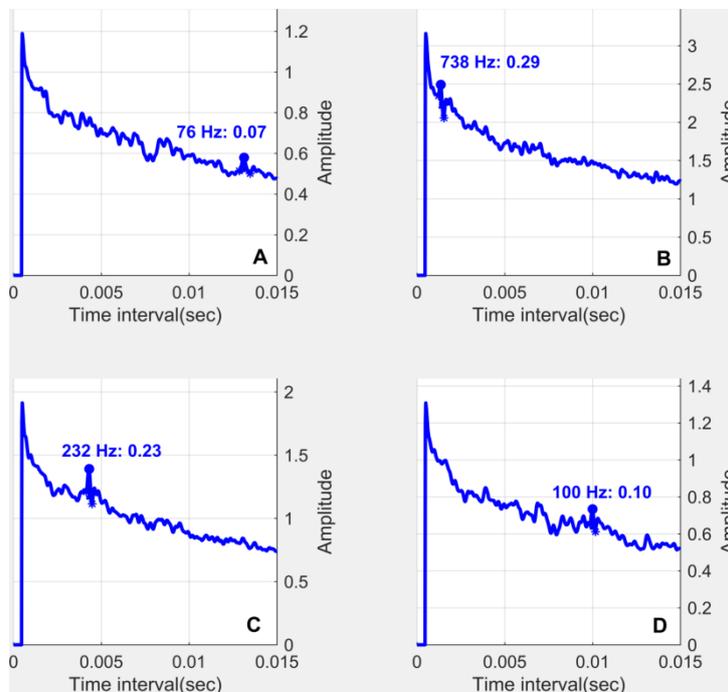

**Fig S.2. The temporal profiles of the stabilized auditory image for a 500 ms signal recorded in the conference room (SN2010 [6])   at 495 ms time frame.** The dot indicates the highest peak and the corresponding values indicate the pitch strength (calculated by Equation 14) and frequency in Hz (using the inverse relationship of time and frequency, f = 1/t).The sub figure A is for the no object recording and sub figures B, C and D are for the recordings with object at 50, 100 and 200 cm respectively.

Initially, the sf2003 module uses an adaptive strobe threshold to issue a strobe on the NAP. After the strobe is initiated, the threshold initially rises along a parabolic path and then returns to the linear decay to avoid spurious strobes (cf. Fig 8). When the strobes  have been computed for each frequency channel of the NAP, the ti2003 module uses the strobes to initiate a temporal integration.

The time interval between the strobe and the NAP value determines the position where the NAP value is entered into the SAI. For example, if a strobe is identified in the 200Hz channel of the NAP at 5 ms time instant, then the level of the NAP sample at 5 ms time instant is added to the 1st position of the 200 Hz channel in the SAI. The next sample of the NAP is added to the 2nd position of the SAI. This process of adding the levels of the NAP samples continues for 35 ms and terminates if no further strobes are identified.

In the case of strobes detected within the 35ms interval, each strobe initiates a temporal integration process. To preserve the shape of the SAI to that of the NAP, ti2003 uses weighting, viz. new strobes are initially weighted high (also the weights are normalized such that the sum of the weights is equal to 1) so that the older strobes contribute relatively less to the SAI. In this way the time axis of the NAP is converted into a time interval axis of the SAI.

The temporal profile in the sub figures of Fig S.2 was generated by summing the SAI along the center frequencies. Fig S.2 shows that the recording with no reflecting object had a pitch strength of 0.07, while the recording with the object at 200 cm (the fourth subfigure in Fig S.2) had a pitch strength of 0.1 at the corresponding frequencies of the repetition pitch. If this is the case for all the recordings has to be verified.

Previous researchers (Yost [26]; Patterson et al [45]) analyzed the perception of repetition pitch by the autocorrelation function. We followed the same approach, since autocorrelation appears to be a good description of how repetition pitch is processed in the auditory system. To determine whether it is autocorrelation or strobe temporal integration that best accounts for human echolocation, repetition pitch and the relevant physiological processes of the auditory system, further analysis is needed, where these two concepts are studied and compared in a number of different conditions.